\documentclass[aps,prb,twocolumn,reprint]{revtex4-1}
\usepackage{graphics}
\usepackage{caption}
\usepackage{subcaption}
\usepackage{epstopdf}
\usepackage{epsfig}
\usepackage{graphicx}
\usepackage{epsf,epic}
\usepackage{color}
\usepackage{amsmath}
\usepackage{booktabs}
\usepackage{multirow}
\usepackage{physics}
\usepackage{hyperref}
\usepackage{amsfonts}
\usepackage{wrapfig}
\usepackage{breqn}
\usepackage{pstricks}
\usepackage[utf8x]{inputenc}
\usepackage{multirow}
\usepackage{fancyref}
\usepackage{pst-node}
\usepackage{float}
\usepackage{bm}
\usepackage{dcolumn}
\newcommand{\etal}{\textit{et al.\ }}

\makeatletter
\usepackage{etoolbox} %
\appto{\appendix}{%
	\@ifstar{\def\theequation@prefix{A.}}%
	{}%
}
\preto\maketitle{%
  \begingroup\lccode`~=`,
  \lowercase{\endgroup
  \let\saved@breqn@active@comma~%
  \let~}\active@comma %
}
\appto\maketitle{%
  \begingroup\lccode`~=`,
  \lowercase{\endgroup
  \let~}\saved@breqn@active@comma %
}
\makeatother
\begin{document}
\title{Electron microscopy and spectroscopic study of structural changes, electronic properties and  conductivity in annealed Li$_x$CoO$_2$}
\author{Halyna Volkova$^a$, Kevin Pachuta$^b$, Kyle Crowley$^c$, Santosh Kumar Radha$^c$, Emily Pentzer$^d$, Xuan P.A. Gao$^c$, Walter R. L. Lambrecht$^c$, Alp Sehirlioglu$^b$,
  Marie-H\'el\`ene Berger$^a$}
\affiliation{$^a$ MINES Paris, PSL University, Centre des Mat\'eriaux, CNRS UMR 7633, BP 87 91003 Evry Cedex, France}
\affiliation{$^b$ Department of Materials Science and Engineering,
  Case Western Reserve University, Cleveland, Ohio 44106-7204, USA}
\affiliation{$^c$ Department of Physics, Case Western Reserve University, Cleveland, Ohio 44106-7079, USA}
\affiliation{$^d$Texas A\&M University, College Station, TX 77843-3003}
\begin{abstract}
   Chemically exfoliated nanoscale few-layer thin Li$_x$CoO$_2$ samples
   are studied as function of annealing at various temperatures,
   using transmission  electron
   microscopy (TEM) and  Electron Energy Loss Spectroscopies (EELS)
   in various energy
  ranges, probing the O-$K$ and Co-$L_{2,3}$ spectra as well as low energy
  interband transitions.  These spectra are compared with first-principles
  density functional theory (DFT) calculations.
  A gradual disordering of the Li and Co cations in the lattice is observed
  starting from a slight distortion of the pure LiCoO$_2$ $R\bar{3}m$
  to  $C2/m$ due to the lower Li content, followed by a $P2/m$ phase
  forming at $\sim$200$^\circ$C indicative of Li-vacancy  ordering,
  formation of a spinel type $Fd\bar{3}m$
  phase around 250$^\circ$C and ultimately a rocksalt type $Fm\bar{3}m$ phase
  above 350$^\circ$C.
  This disordering leads to a lowering of the band gap
  as established by low energy EELS. The Co-$L_{2,3}$
  spectra indicate a change of average Co-valence from an initial value of about 3.5 consistent with Li-deficiency related Co$^{4+}$,
  down to 2.8 and 2.4 in the $Fd\bar{3}m$ and $Fm\bar{3}m$, indicative of
  the increasing presence of Co$^{2+}$ in the higher temperature
  phases. The O-$K$ spectra  of the rocksalt phase are only
  reproduced by a calculation for pure CoO and  not for a model with random
  distribution of Li and Co. This indicates that there may be a  loss of Li
  from the rocksalt regions of the sample at these higher temperatures. The
  conductivity measurements indicate a gradual drop in conductivity above 200$^\circ$C. This loss in conductivity is clearly related to the more Li-Co interdiffused phases, in which a low-spin electronic structure is no longer valid
  and stronger correlation effects are expected. Calculations for these phases
  are based on DFT+U with Hubbard-U terms with a random distribution of magnetic moment orientations, which lead to a gap even in the paramagnetic phase
  of CoO. 
\end{abstract}
\maketitle
\section{Introduction}
The discovery of graphene and its associated ultrahigh electron mobility\cite{Novoselov04,Rahimi2014} led to interest in fundamental and applied research
on other two-dimensional (2D) atomically thin materials,
such as transition metal dichalcogenides,\cite{Mak10,Radisavljevic2011}
black phosphorus,\cite{Li2014} antimonene.\cite{Ares18}
2D oxides have received relatively
less attention thus far,\cite{Yang19}
although many oxides exist in layered forms
which might be amenable to exfoliation. Furthermore, the 
formation of a two-dimensional electron gas (2DEG), the prime playing ground
for high-electron mobility, has been found possible at  oxide surfaces
and interfaces.\cite{Altfeder16,Wang14,Gonzalez17,Radha2020}
Two dimensional oxides may also be expected
to be more stable in extreme environments and have a multitude of
functionalities, from catalysis to varying types of magnetic and
ferro-electric ordering, superconductivity  and metal-insulator transitions.\cite{Marianetti2004,Nguen19,Takada2003,Iwaya13}
However, their potential for high-speed
ultrathin transistors has not yet been established. While some 2D oxides,
such as MoO$_3$ and V$_2$O$_5$ exhibit van der Waals bonded neutral
layers and can be mechanically exfoliated,\cite{Sucharitakul2017,Zhang2017}
they require doping or intercalation to increase the conductivity. 
Other oxides such as LiCoO$_2$ exhibit a layering of  the
two cations with alternating CoO$_2^{-1}$  and Li$^{+1}$
layers which are electrostatically bonded. Such materials can be exfoliated by a combination of
chemical and mechanical exfoliation techniques to achieve nanosize few
atomic layer thin flakes. While the properties of bulk Li$_x$CoO$_2$
as function of Li concentration have been amply studied in the context of
LiCoO$_2$ based batteries,\cite{Iwaya13,Miyoshi18,Wu2014}
the properties of nanoflakes are still not well known.

Recently, a chemical exfoliation procedure was
established for LiCoO$_2$.\cite{Pachuta,Pachuta20,Masuda06}
Using additional mechanical
exfoliation it becomes possible to also study the conductive properties
of such ultrathin nanoscale samples.\cite{Crowley2020} However, preparing contacts to
these nanoscale structures requires annealing and the properties of
these materials as function of temperature therefore need to be
understood. This article focuses on LiCoO$_2$ nanoflakes 
with the study of their crystallographic structure, electronic structure and conductive properties. The electronic structure is probed using various forms
of Electron Energy Loss Spectroscopy (EELS)  and the structure and morphology
are studied using Selected Area Diffraction (SAED) in a transmission electron
microscope (TEM).
The spectra are interpreted with various types of first-principles
electronic structure calculations and trends are established as function of
temperature and correlated with the structural evolution of the samples. 

\section{Experimental Methods}
The Li$_{0.37}$CoO$_2$ nanoflakes were obtained by exfoliation in three
steps.\cite{} First, Li$^+$  was substituted by H$^+$ in a 1M HCl
aqueous solution. Then, H$^+$ was replaced by larger molecules of
tetramethylammonium hydroxide (N(CH$_3$)$_4$-OH, TMAOH) to expand 
the space between CoO$_2$ layers. The H associates with the OH group
and leaves the same as H$_2$O while the TMA$^+$ ions replace the initial
Li$^+$.  Due to the increased interlayer distance, the 
CoO$_2^{-1}$  nanosheets then enter the solution but will there still
associate with positive ions, which depend on which salts are present in the
solution. 
Finally, the nanosheets were re-precipitated by adding Li$^+$
ions to the solution in the form of LiCl. Other salts, such as NaCl, KCl
etc. can be used in this step but here we use Li exclusively. 
The nanoflakes of Li$_x$CoO$_2$ had thicknesses, ranging from 30 to 160 nm
as measured by Atomic Force Microscopy (AFM) and
Electron Energy Loss Spectroscopy (EELS). Note that the conventional cell of
$R\bar{3}m$ has a hexagonal $c$-axis of $\sim$1.4 nm and contains
three CoO$_2$ layers so that the thinnest flakes are still
about 60 CoO$_2$ layers thick. 
The composition Li$_{0.37}$CoO$_2$  was determined by
inductively coupled plasma mass spectroscopy (ICPMS). The Li$_{0.37}$CoO$_2$ nanoflakes
were deposited in a drop of aqueous chemical solution on top of a
degenerately doped Si wafer with a 300 nm SiO$_2$ surface layer.
There were two batches: one for electronic transport properties, and another one for Transmission Electron Microscopy (TEM) and EELS.
For conductivity measurements, individual devices based on Li$_{0.37}$CoO$_2$
nanoflakes were obtained by sputtering nickel contacts ($\sim$ 90 nm thick) and electron beam lithography. The current-voltage response was measured
on $\sim$45-160 nm nanoflakes, using a Physical Property Measurement System (Quantum Design Inc.).
High-temperature annealing at 150, 200, 250, 300, and 350°C for $\sim$30 minutes were performed in vacuum, followed by a gradual cooling ($< 10$ K/min).

The ex-situ TEM was performed by transferring annealed nanoflakes onto TEM holey carbon films supported by Cu grids.
The FEG-TEM analysis was performed on a Tecnai F20ST (FEI) operating at 200 kV
and equipped with GIF 2000, Gatan, for Energy Filtered TEM (EFTEM) and parallel EELS.
The Selected Area Electron Diffraction (SAED), Bright field (BF)/Dark Field (DF) images, energy-filtered HRTEM, and EELS spectra were acquired from individual
flakes. The spectra were acquired in diffraction mode,
with a camera length of 62 mm and a GIF entrance aperture of 0.6 mm
leading to collection semi-angle of 2.8 mrad. The dispersion was set to 0.1 eV/channel, and a FWHM of the zero loss electrons peak of $\sim$0.6 eV was obtained. The spectra were acquired, as close to the $(000)$ transmitted beam as
possible,  with the probe radius of 1.15 nm-1 in the horizontal section of reciprocal space, which is represented by a TEM diffraction pattern. This probe radius is $~\sim2-4$ times smaller than the $1/d_{hkl}$ reciprocal space distances from $(000)$ to neighboring Brillouin zone centers (neighboring diffraction reflections).
This allowed to probe
electron transitions with ${\bf q}\approx0$ momentum transfer. Essentially
one measures the longitudinal ${\bf q}=0$ response of the system
in this manner, which is theoretically represented by $-\mathrm{Im}[\varepsilon^{-1}({\bf q}\approx0,\omega)]$. 
The spectra were corrected for background and multiple scattering events, and low-loss spectra were deconvoluted by zero-loss (ZL) peak recorded in vacuum, using softwares Digital Micrograph and PEELS (ref P.Fallon, C.A.  Walsh, PEELS Program,University of Cambridge, UK, 1996). Fityk was used for fits. The VESTA software\cite{VESTA} was used for crystal structure visualization.
\section{Computational Methods}\label{sec:theory}
The underlying approach for our calculations is density functional
 theory (DFT) in the 
 generalized gradient approximation (GGA) in the Perdew-Burke-Ernzerhof (PBE) parametrization\cite{PBE}.
 Calculations of the band structure and partial densities of states
 were done using the full-potential linearized
 muffin-tin-orbital (FP-LMTO) method
 as implemented in questaal.\cite{questaal,questaal-paper}
  Convergence parameters for the LMTO calculations were chosen
as follows: basis set $spdf-spd$ spherical wave envelope functions  plus augmented plane waves with a cut-off of 3 Ry, augmentation cutoff $l_{max}=4$,
{\bf k}-point mesh, $12\times12\times12$.
We used experimental lattice constants but relaxed the internal
structural parameters.

Because some of the materials considered, such as CoO and Co$_3$O$_4$ are
incorrectly found to be metals within GGA,
we also use the DFT+U approach in which
on-site Coulomb terms  are added for the Co-$3d$ orbitals
as Hubbard-U parameters. We used the same  value of $U=5$ eV as in
Trimarchi\etal.\cite{Trimarchi18}
In LiCoO$_2$ in the $R\bar{3}m$ structure, 
this shifts the already empty $d$-$e_g$ band  up relative to the filled
bands but keeps the low-spin configuration. In CoO and Co$_3$O$_4$ this
splits up and down spin orbitals and creates a local magnetic moment.
In Co$_3$O$_4$ we just use the primitive cell and obtain a ferromagnetic
solution. In paramagnetic CoO, he situation is more subtle. We use a special
quasirandom structure (SQS) approach\cite{SQS90,Trimarchi18,YuboZhang20}
in a 64 atom cubic supercell, in
which up and down spin Co sites are randomly placed. More precisely
they are placed such that the pair-correlations between these sites
mimic the fully random ones. This is called a  polymorphous  description
of the magnetic moment disorder. Unlike the approach of Trimarchi \etal\cite{Trimarchi18} we here do not displace the atoms in random local fluctuations.
Our solution may thus be somewhat different but is similar in spirit.
We do not claim here that such fluctuations in position do not occur
in response to the random spin directions. Our main goals is 
to have a qualitatively correct description of the electronic structure
with a gap induced by magnetic moment formation. 

To describe the O-$K$ spectra we use the selection rule that for
small momentum transfer, transitions from the O-$1s$ core state
to the empty bands are essentially proportional to the O-$p$ like partial
density of states (PDOS). One should note, however, that for high energy
states, this does not necessarily correspond to the atomic O-$2p$ states
but rather higher excited atomic states or merely
the tails of surrounding atom wave functions expanded in spherical harmonics
inside the O-sphere. It is well known
that the O-$1s$ core hole presence may modify the higher state O-$p$ like
PDOS and hence we include the core-hole explicitly in the calculation.
This is known as the final state rule in X-ray absorption spectroscopy
but the same applies in EELS.
Core holes are created in a randomly chosen oxygen site in a supercell, 
after which the total electron density is iterated to self consistency.
Using this new self-consistent wave function, the spectrum is calculated 
including the effect of the momentum or dipole matrix elements, 
\begin{equation}
  S(\epsilon)\propto\sum_{n{\bf k}} |\left\langle\psi_{c}|\mathbf{p}| \psi_{n{\bf k}}\right\rangle|^2(1-f_{n{\bf k}})\delta(\epsilon-\epsilon_{n{\bf k}}+\epsilon_{c}), \label{eqOKsim}
\end{equation}
with $\epsilon_{n{\bf k}}$, $\psi_{n{\bf k}}$ the band Kohn-Sham eigenvalue
and the corresponding wavefunction, $\epsilon_c$, $\psi_c$ the core energy
and wave function, 
and $f_{n{\bf k}}$ the Fermi-Dirac occupation factor.
Because we do not attempt to calculate the absolute core-level spectrum edge,
the $\epsilon_c$ is not actually calculated and set to zero and the first
peak is aligned with the experimental spectra. 
For the $R\bar{3}m$ structure a $3\times3\times3$ supercell is used. Similar
size supercells are used in the other cases. A Gaussian broadening of $\sim$1 eV is applied to these spectra to represent instrumental broadening for
easier comparison with the experiment. We have checked that including the
core-hole is not crucial to represent the shape of the spectra and even
a simple $p$-PDOS gives actually similar results. Thus we use the latter
for the SQS representing the rocksalt paramagnetic CoO for which it
was difficult to find a converged solution using the FP-LMTO code.
Instead we used the GPAW code with an LCAO basis set which is also
used for the EELS, as described next. 

The measured low energy EELS are compared with  calculations of
the loss-function $-\mathrm{Im}[\varepsilon^{-1}({\bf q},\omega)$
at small finite {\bf q}. These calculations are performed 
within the generalized random phase approximation (RPA) 
using GPAW\cite{GPAW,Mortensen05,Enkovaara2010} which uses the projector augmented wave (PAW) method \cite{PAW}. The kinetic energy cut-off for the plane wave basis set is taken to be 700 eV with a \textbf{k}-point grid of $20\times20\times20$. The band structures obtained with this method were checked to be
in excellent agreement with the all-electron FP-LMTO results.

To calculate the loss function, we start from  the non-interacting
charge-charge response function ($\chi_{\mathbf{G} \mathbf{G}^{\prime}}^{0}(\mathbf{q}, \omega)$) obtained from the Adler-Wiser formula \cite{adler,wiser}
\begin{eqnarray}
  \chi_{\mathbf{G} \mathbf{G}^{\prime}}^{0}(\mathbf{q}, \omega)&=&\frac{1}{\Omega}
  \sum_{\mathbf{k}} \sum_{n, n^{\prime}} \frac{f_{n \mathbf{k}}-f_{n^{\prime} \mathbf{k}+\mathbf{q}}}{\omega+\epsilon_{n \mathbf{k}}-\epsilon_{n^{\prime} \mathbf{k}+\mathbf{q}}+i \eta} \nonumber \\
  &&\left\langle\psi_{n \mathbf{k}}\left|e^{-i(\mathbf{q}+\mathbf{G}) \cdot \mathbf{r}}\right| \psi_{n^{\prime} \mathbf{k}+\mathbf{q}}\right\rangle_{\Omega_{\mathrm{cell}}}\nonumber \\
    &&\left\langle\psi_{n \mathbf{k}}\left|e^{i\left(\mathbf{q}+\mathbf{G}^{\prime}\right) \cdot \mathbf{r}^{\prime}}\right| \psi_{n^{\prime} \mathbf{k}+\mathbf{q}}\right\rangle_{\Omega_{\mathrm{cell}}},  \label{eq:1}
\end{eqnarray}
where $\Omega$ is the crystal volume. 
Here, \textbf{G} and \textbf{q} are the reciprocal lattice vector and the
wave vector of the charge density perturbation respectively.

Within Time Dependent Density Functional Theory (TDDFT), one can write the interacting charge-charge response function by solving a Dyson equation of the form
\begin{equation}
	\begin{aligned}
	\chi_{\mathbf{G G}^{\prime}}(\mathbf{q}, \omega)=&\chi_{\mathbf{G G}^{\prime}}^{0}(\mathbf{q}, \omega)+\\ &\sum_{\mathbf{G}_{1} \mathbf{G}_{2}} \chi_{\mathbf{G G}_{1}}^{0}(\mathbf{q},\omega) K_{\mathbf{G}_{1} \mathbf{G}_{2}}(\mathbf{q}) \chi_{\mathbf{G}_{2} \mathbf{G}^{\prime}}(\mathbf{q}, \omega). \label{eq:2}
	\end{aligned}
\end{equation}
Here the kernel $K_{\mathbf{G}_{1} \mathbf{G}_{2}}$  is treated in the
random phase approximation (RPA) and hence only includes the Coulomb part. 
\begin{equation}
K_{\mathbf{G}_{1}\mathbf{G}_{2}}(\textbf{q})=\frac{4\pi}{\abs{\textbf{q}+\textbf{G}_1}^2}\delta_{\textbf{G}_1 \textbf{G}_2} \label{eq:3}
\end{equation}
Using $\chi_{\mathbf{G} \mathbf{G}^{\prime}}(\mathbf{q}, \omega)$, the macroscopic dielectric function $\epsilon_M$ is obtained to be
\begin{equation}
  \varepsilon_{M}^{-1}(\mathbf{q}, \omega)=1+\frac{4\pi}{|{\bf q}+{\bf G}|^2}
  \chi_{\mathbf{G}=\mathbf{0},\mathbf{G}^{\prime}=\mathbf{0}}(\mathbf{q}, \omega)\label{eq:4}
\end{equation}

The ${\bf q}$ dependent dynamical loss function can be directly related to the loss function  which is calculated as 
\begin{equation}
\mathcal{A}_{\operatorname{EELS}}(\mathbf{q}, \omega)=-\text{Im}\left[\varepsilon_{M}^{-1}({\bf q}, \omega)\right]
\end{equation}

In addition, we may neglect local field effects (LFE), in which
case 
\begin{equation}
  \varepsilon^{NLFE}_{{\bf G},{\bf G}^\prime} ({\bf q},\omega)=\delta_{{\bf G},{\bf G}^\prime}
-\frac{4\pi}{|{\bf q}+{\bf G}|^2}\chi^0_{{\bf G},{\bf G}^\prime}({\bf q},\omega)
\end{equation}

and
\begin{equation}
  \varepsilon_M^{-1}({\bf q},\omega)\approx 1/\varepsilon^{NLFE}_{{\bf G}={\bf 0},{\bf G}^\prime={\bf 0}}({\bf q},\omega)]
\end{equation}
\section{Results}
\subsection{TEM diffraction observations}

\begin{figure*}
  \begin{subfigure}[b]{\textwidth}
    \includegraphics[width=\textwidth]{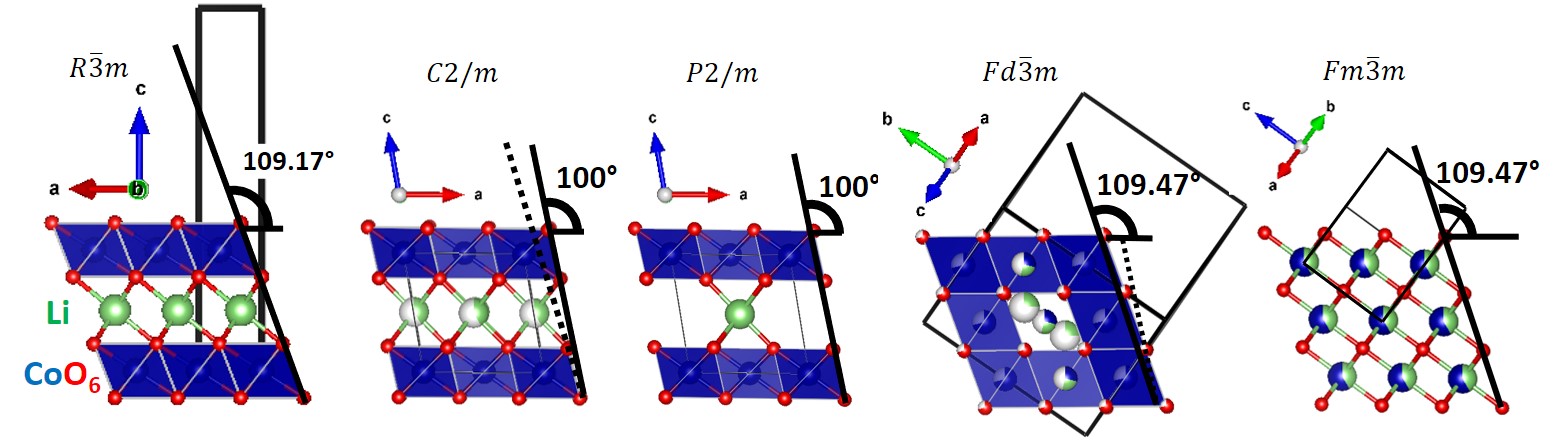}
    \caption{Schematic crystal structures.\label{figstrucs}}
  \end{subfigure}
  \begin{subfigure}[b]{0.15\textwidth}
    \includegraphics[width=\textwidth]{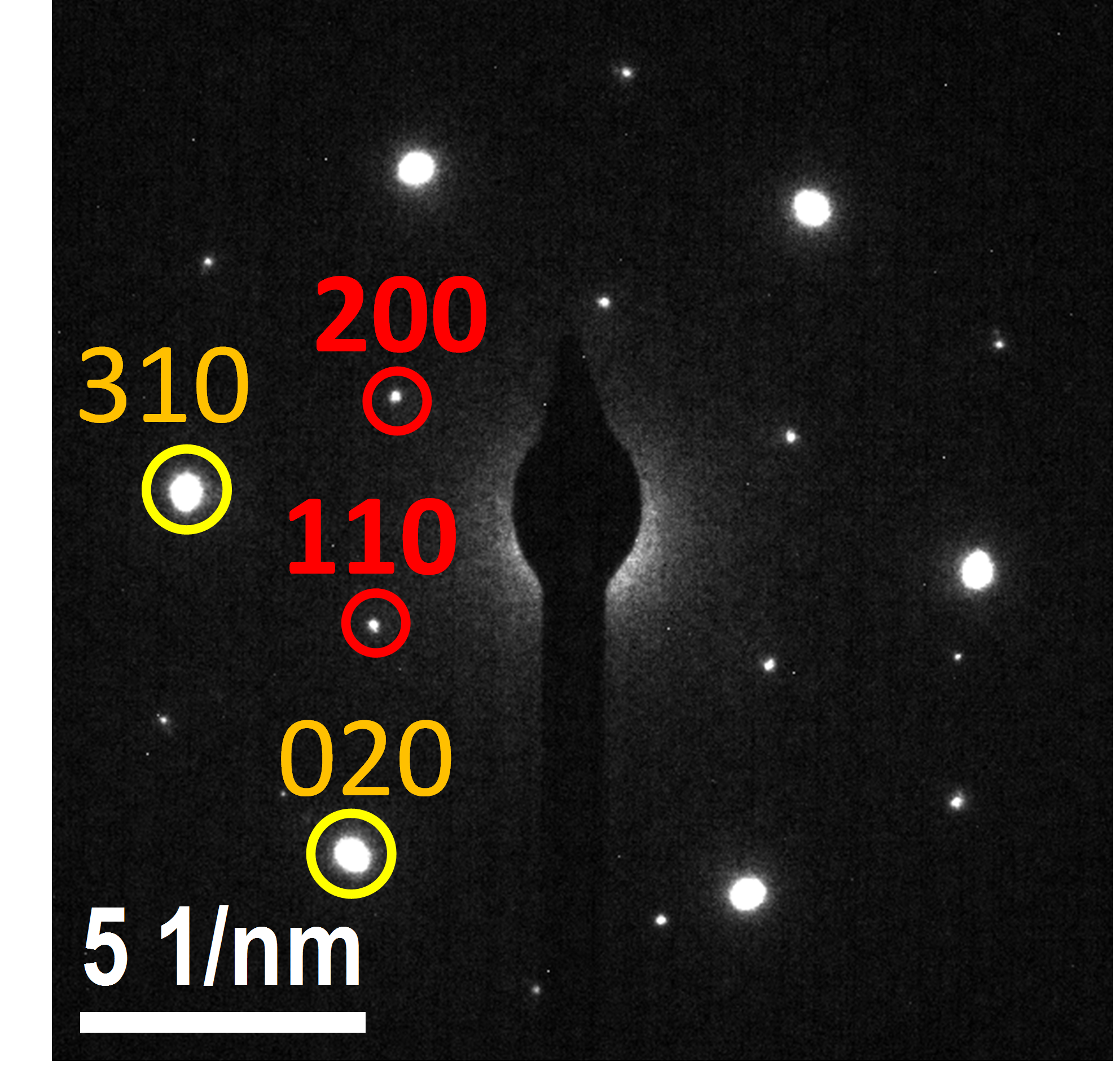}
    \caption{20$^\circ$C $R\bar{3}m+C2/m$ $\langle001\rangle$.\label{saed20}}
  \end{subfigure}
  \begin{subfigure}[b]{0.15\textwidth}
    \includegraphics[width=\textwidth]{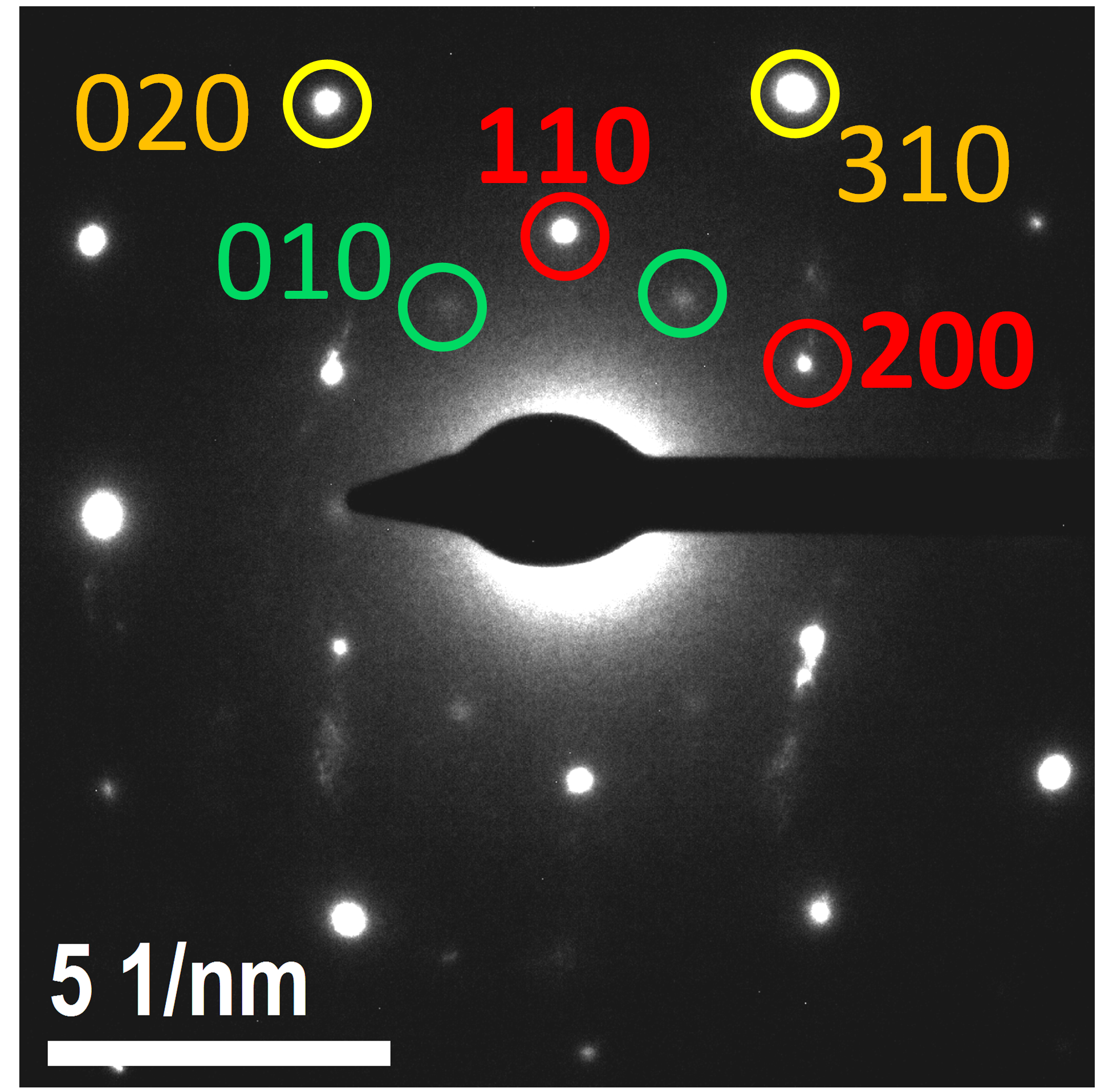}
    \caption{150$^\circ$C $R\bar{3}m+C2/m$ $\langle001\rangle$.\label{saed150}}
  \end{subfigure}
  \begin{subfigure}[b]{0.15\textwidth}
    \includegraphics[width=\textwidth]{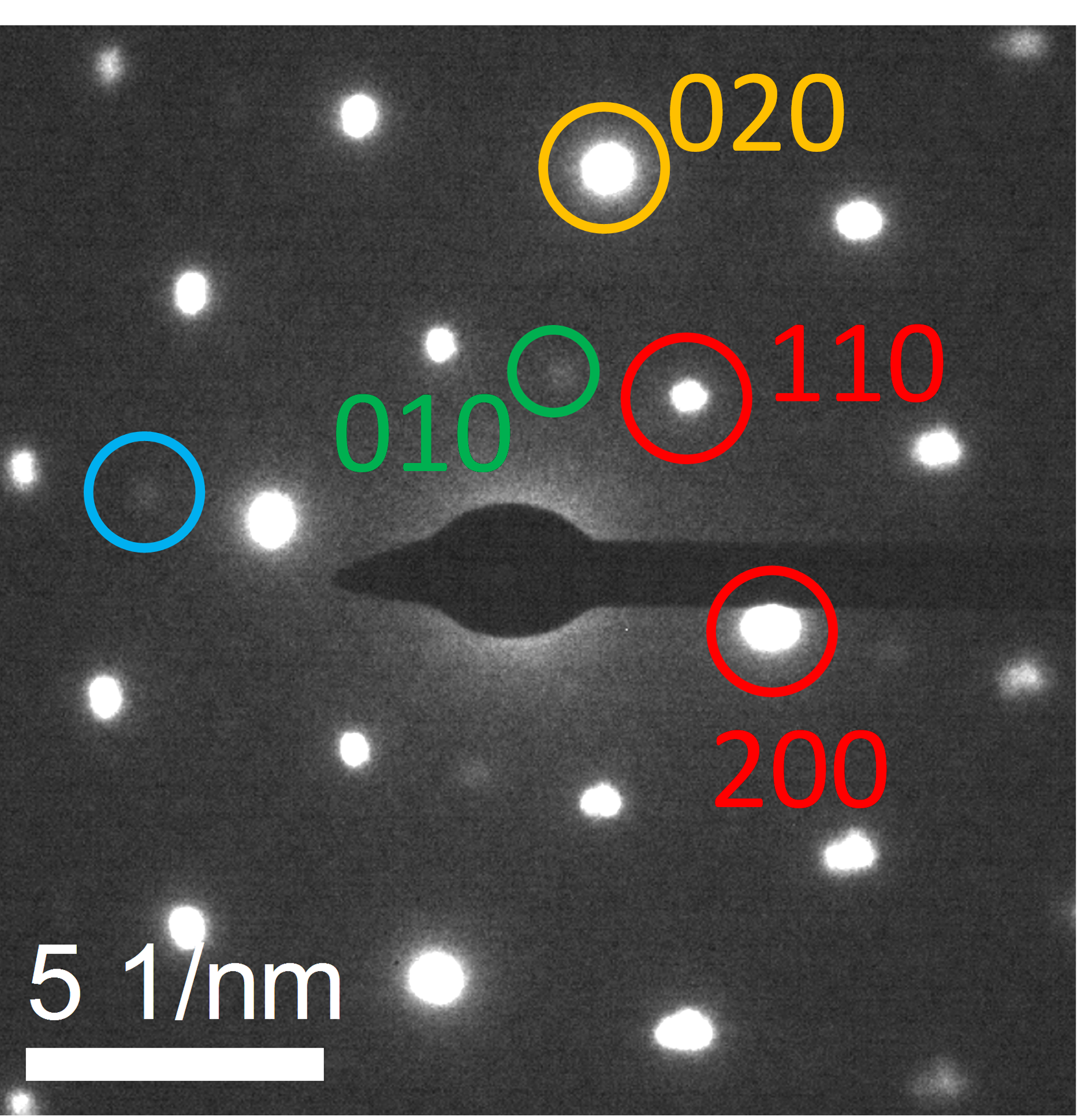}
    \caption{200$^\circ$C \\$R\bar{3}m$+$C2/m$+$P2/m$ \\$\langle001\rangle$.\label{saed200}}
  \end{subfigure}
  \begin{subfigure}[b]{0.15\textwidth}
    \includegraphics[width=\textwidth]{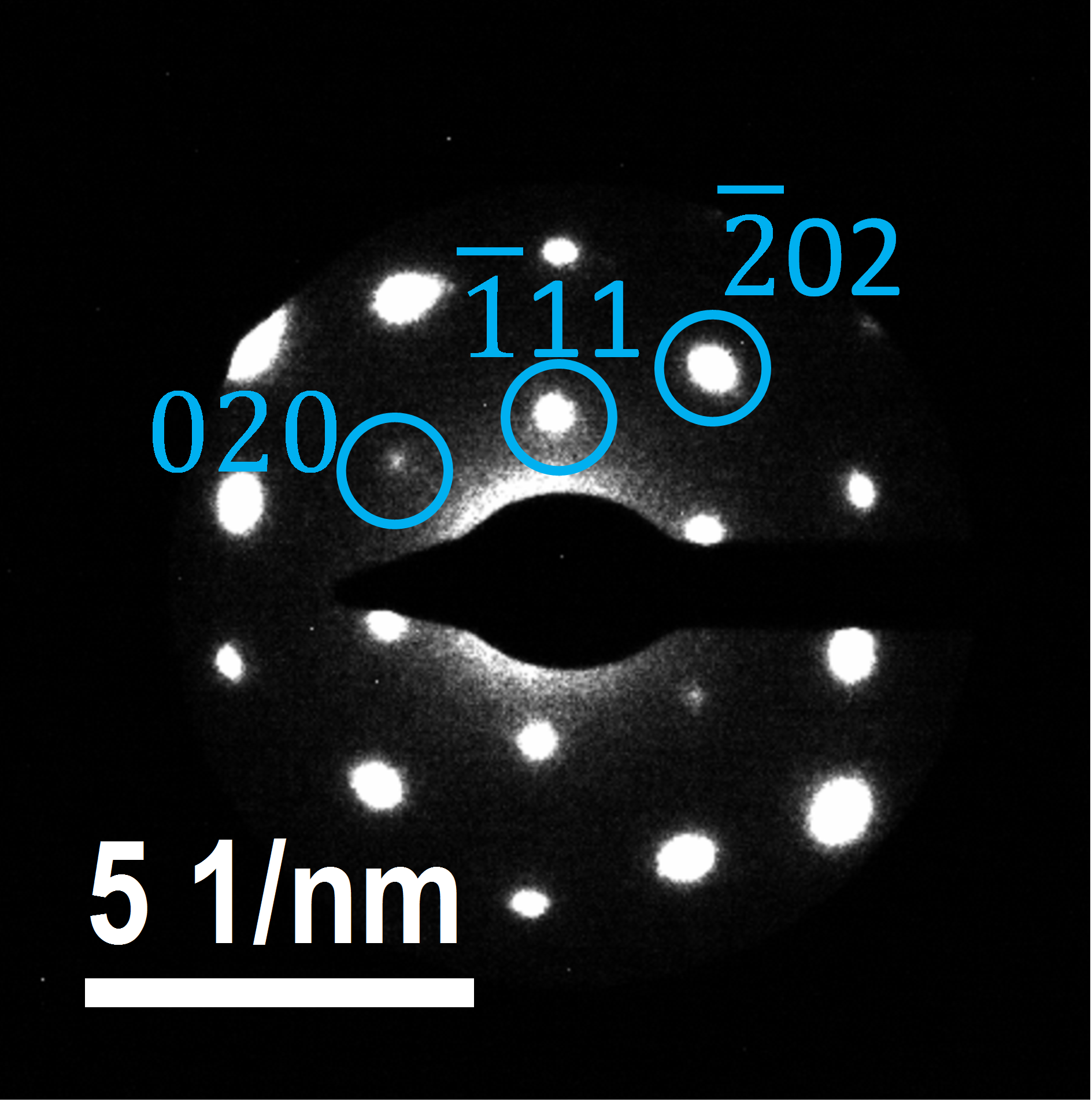}
    \caption{250$^\circ$C \\$Fd\bar{3}{m}$ \\$\langle101\rangle$.\label{saed250}}
  \end{subfigure}
\begin{subfigure}[b]{0.15\textwidth}
\includegraphics[width=\textwidth]{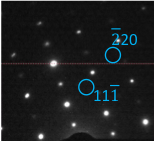}
\caption{250$^\circ$C \\$Fd\bar{3}{m}$ \\$\langle112\rangle$.\label{saed2502}}
  \end{subfigure}
  \begin{subfigure}[b]{0.15\textwidth}
    \includegraphics[width=\textwidth]{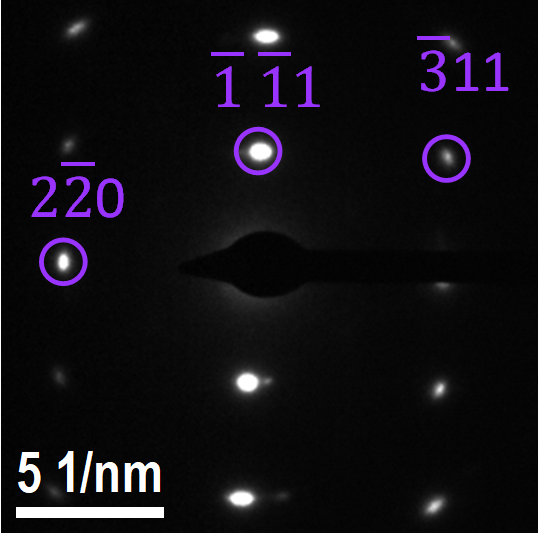}
    \caption{350$^\circ$C\\ $Fm\bar{3}{m}$\\ $\langle112\rangle$.\label{saed350}}
  \end{subfigure}
  \caption{Structure evolution upon annealing: (a) the schematic shows a repeatable part of structure which is preserved on phase transformations (not a unit cell). The pure CoO$_6$ atomic columns consist of blue octahedra. (b-g)SAED patterns for different annealing temperatures. \label{figsaed}}
\end{figure*}
Fig. \ref{figsaed} illustrates the structure evolution of the Li$_x$CoO$_2$
nanoflakes with  temperature, determined through the changes of
TEM SAED patterns.
The nanosheets typically were found lying with their octahedral sheets parallel to the carbon film. Before annealing, the diffraction pattern of a flake
(Fig. \ref{saed20}) showed two arrays of spots, an intense one circled
in yellow that corresponds to a single crystal with $R\bar{3}m$ symmetry
observed in the $[0001]$ zone axis, and a fainter array of spots circled in red,
that corresponds to the formation of coherent nano-domains with the $C2/m$
symmetry. The latter is due to the slight distortion of the cell due to the
loss of lithium (Fig. \ref{figstrucs}) and the resulting repulsion
between the CoO$_2$ layers. This structure corresponds to a random
distribution of vacancies on the Li sites. It was shown before \cite{Motohashi09,Chang13}
that the ``O3''-type $C2/m$ phase (Fig. \ref{figstrucs}) was stabilized for
$x\approx0.2-0.5$. A unique crystal orientation relationship
with the $R\bar{3}m$ matrix was found, described as $[0001]\parallel[001]$,
$(11\bar{2}0)\parallel(020)$.
For more structural details, see Supplementary Table ST1.\cite{SM}

After annealing at 150$^\circ$C, the monoclinic $C2/m$ domains were
coarsening reaching the size of the former $R\bar{3}m$ domains
(Supplementary Fig. S2),\cite{SM} as also evidenced in Fig. \ref{figsaed}c
through growing intensity of the corresponding reflections (in red).
The prolonged TEM observation favored the appearance of additional 010 spots
(in green), forbidden by the C lattice symmetry.
These additional spots evidence the beginning of the phase transformation into
the $P2/m$ phase on beam damage.
In the $P2/m$ phase, 1:1 in-plane ordering of the Li-vacancy in
Li$_{0.5}$CoO$_2$ takes place\cite{Kang2019} (Fig. \ref{figstrucs}),
as compared to the $C2/m$ phase. In Fig. \ref{saed150}, the
three pairs of new reflections are attributed to three domain orientations
\cite{Shao_HOrn_2003}, which we also observed through energy filtered HRTEM (Supplementary Fig. S1(2)).\cite{SM}
For $T=200^\circ$C (Fig. \ref{saed200}), the $C2/m$ and $P2/m$
phases coexist. The phase  transformation of monoclinic phase into
a cubic spinel-type $Fd\bar{3}m$ forming nano-domains was seen in HRTEM
(Supplementary figs. S1(3,4).\cite{SM}
The orientation relationship (Figs. \ref{figsaed}d-e)
was found to be  $[010]_P\parallel[101]_F$ and $[001]_P\parallel\approx [11\bar{1}]_F$.   

The phase transformation to spinel was confirmed after annealing at $250^\circ$C (Fig. \ref{figsaed}d-e, where the
$Fd\bar{3}m$ phase was macroscopic (Fig. \ref{figsaed}d-e and
Supplementary Fig.S1).\cite{SM}
The transformation results from inter-layer diffusion of Co/Li, with octahedral and tetrahedral site occupancies. Combining our SAED with characteristic
spot intensities (Fig. \ref{saed250}), the EFTEM rod-like contrast
(Supplementary Fig. S1(4)),\cite{SM} and simulations of SAED/EFTEM of known
spinel phases\cite{Wang_2019,Gummov93},
we proposed the structure shown in Fig. \ref{figstrucs}, with  8a tetragonal
sites partially occupied by Li and 16d (blue octahedra)
and 16c sites alternating in the non-diagonal columns occupied by Co.
The simulation of
diffraction patterns were made in the $\langle211\rangle$
zone axis, since it allowed a better discrimination of the spinel
types by the relative intensities of their diffraction spots than in the
$\langle111\rangle$ zone axis (Supplementary Fig. S3a).\cite{SM}
The EFTEM contrast was simulated in $[101]$ zone axis
(Supplementary Fig. S1(4-5)).\cite{SM} We propose the suitable atomic site occupations, which are given in Supplementary table ST1.\cite{SM} More studies are necessary to determine the precise site occupation. 
For $T=300^\circ$C, the flakes stayed in
the $Fd\bar{3}m$ phase (Supplementary figs. S1-5 and  S2-b).\cite{SM}

For $T= 350^\circ$C, some of the flakes transformed into the rocksalt
$Fm\bar{3}m$ phase (Fig.\ref{figstrucs}) with
about  half lattice parameter (4.2 \AA)
and the same orientation of the $a$, $b$, $c$ lattice vectors. The transition is
due to full mixing of Co and Li in 4a octahedral sites of $Fm\bar{3}m$
(fig. 1a) corresponding to the former 16c and 16d sites of the
$Fd\bar{3}m$ structure. The partial phase transition is explained by
different flake thicknesses, since it controls the rate of oxygen loss
\cite{Sharifi-Asl2017}. On top of newly formed $Fm\bar{3}m$, a remnant $Fd\bar{3}m$ phase
was observed, which was similar to a fully disordered spinel\cite{Wang_2019}.
The energy filtered HRTEM (Supplementary fig.S1-6)\cite{SM} showed the contrast, typical for spinel-like structure with the 8.3 \AA\ lattice parameter, but with empty 8a sites (Supplementary fig. S3c).  

\subsection{Overview of electronic transitions}
\begin{figure}
  \includegraphics[width=8cm]{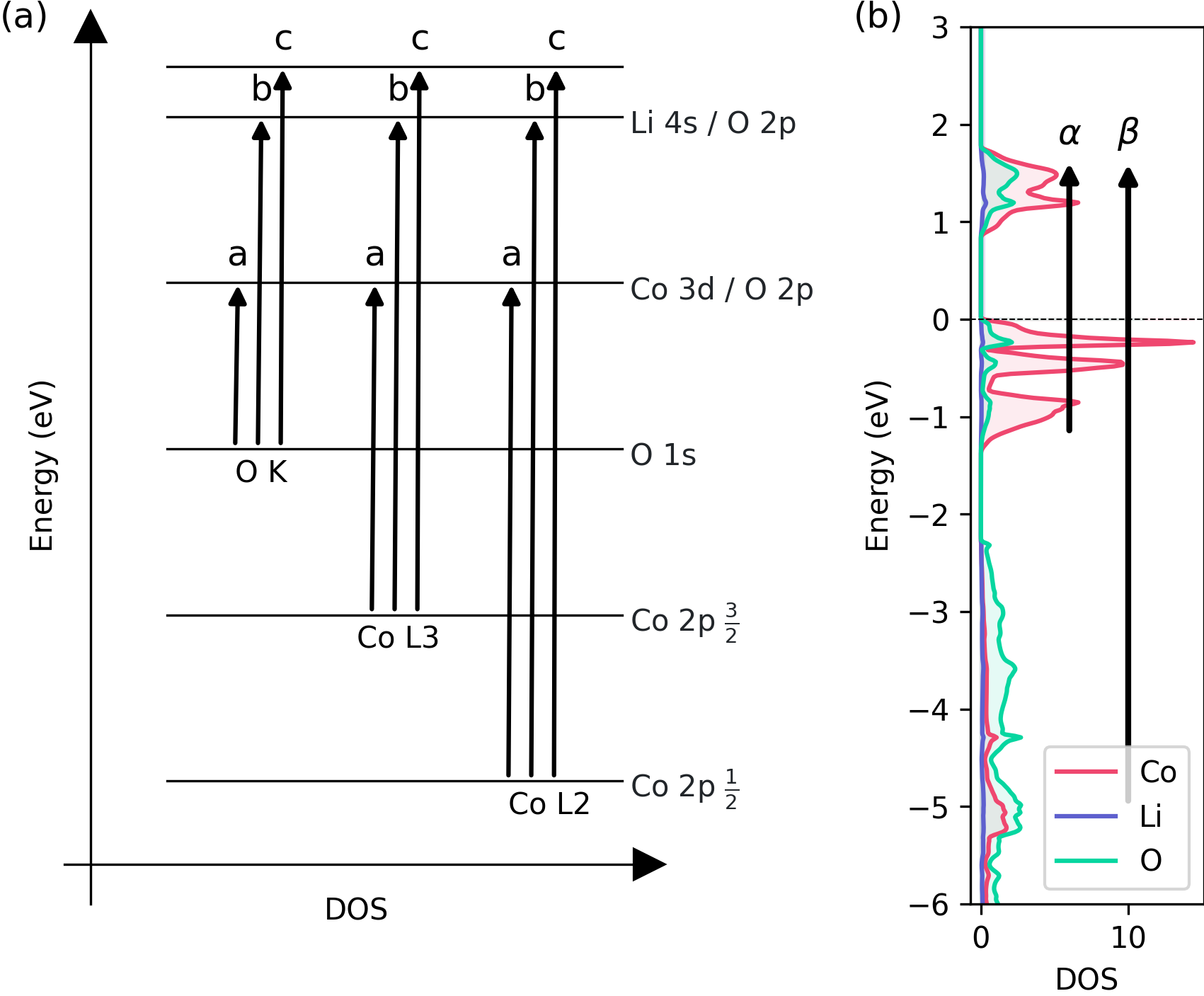}
  \caption{Schematic of energy transitions.\label{figschematic}}
\end{figure}

Electronic structure changes induced by the structural modifications are studied in the following parts through EELS spectra of O-$K$ and Co-L edges.  Fluctuations near the absorption edges are related to the probability of electron transitions from occupied to unoccupied electron levels. A schematic overview
of the electronic transitions discussed here is given in Fig.\ref{figschematic}

\begin{figure}
  \includegraphics[width=8cm]{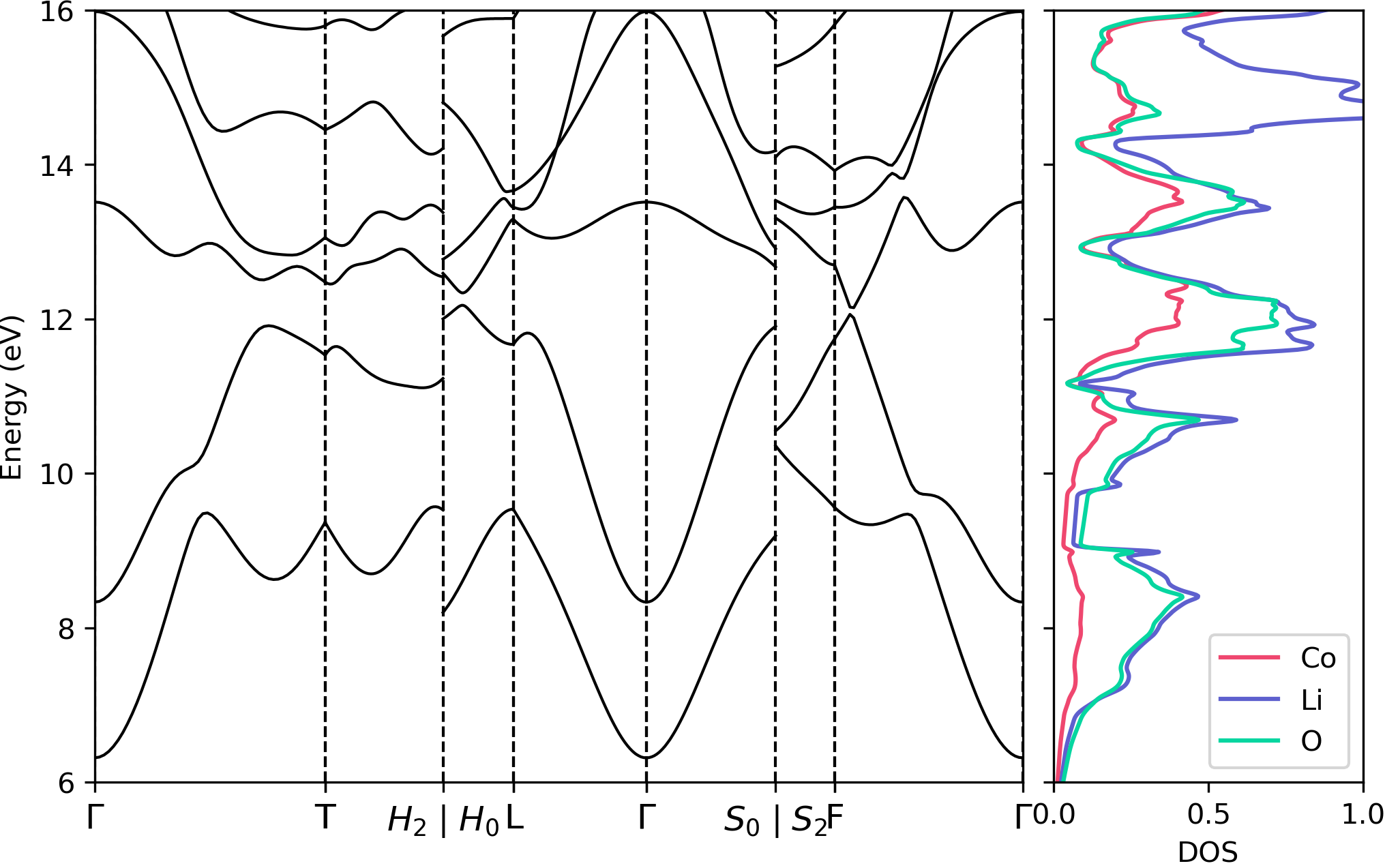}
  \includegraphics[width=8cm]{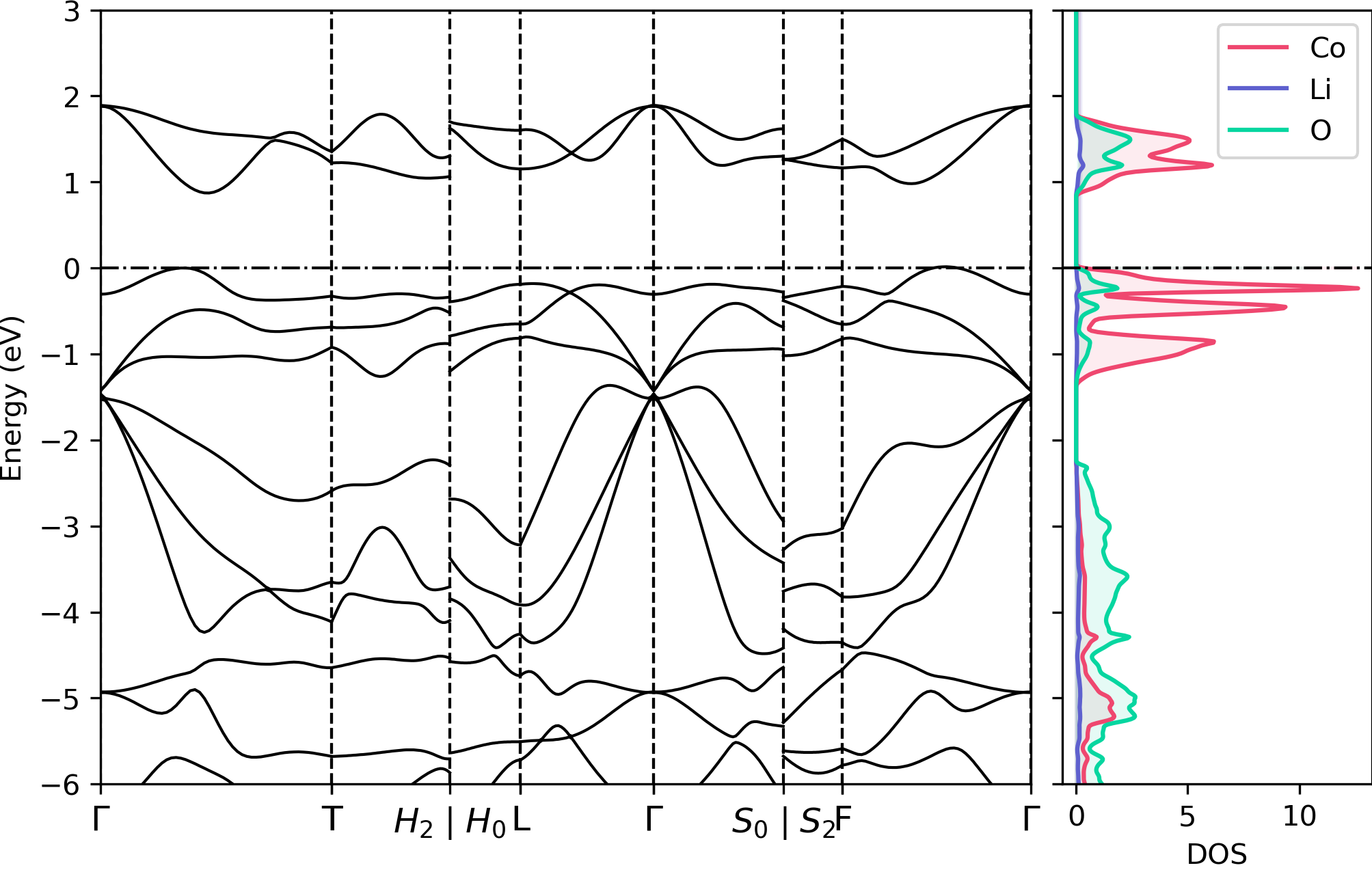}
  \caption{GGA band structure and PDOS of LiCoO$_2$ in $R\bar{3}m$ structure
    in two energy intervals.
  \label{lda-bands-dos}}
\end{figure}

The GGA band structure and DOS of LiCoO$_2$ in the $R\bar{3}m$ structure
are shown in Fig. \ref{lda-bands-dos}. Please note the different PDOS scale
in the higher lying conduction band region. The top of the valence band
is Co-$d$-$t_{2g}$ like, while deeper bands have more O-$2p$ character.
The lowest conduction band is dominated by Co-$3d$ while the higher bands in the range 6-10 eV have mainly
Li-$2s2p$ and O-antibonding contributions. There is also a Co-$4s$ contribution
which however is stronger between  12-14 eV. 
The bands in this energy range have strong dispersion
and are essentially free-electron like.
In the GGA, the band gap is somewhat underestimated and amounts
to 0.89 eV for the indirect gap. 

\subsection{O-$K$ edge spectra}
\begin{figure}
  \includegraphics[width=8cm]{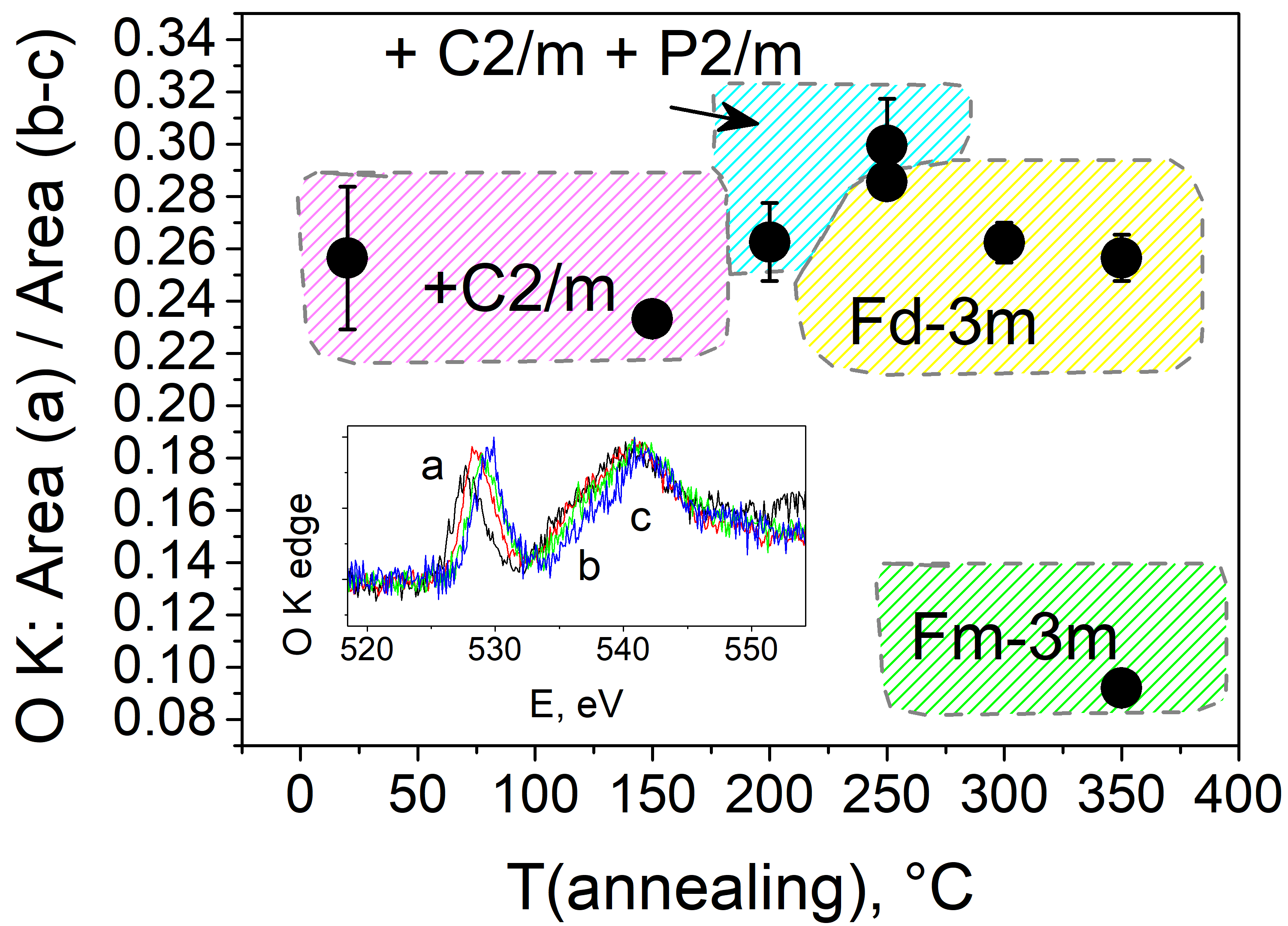}
  \caption{EELS study of O K edge: (a) the ratio of integrated area $A(a) / A(b,c)$ of the pre-peak ``a'' versus main peak components ``b'' and ``c'',
    as identified 
    in the inset, which  shows the spectrum of Li$_{0.37}$CoO$_2$
    at 20$^\circ$~C. \label{figOK}}
    \end{figure}

In Fig. \ref{figOK}, the inset shows the typical EELS O-$K$ edge spectrum
for 20$^\circ$C (non-annealed) sample (spectra recorded at all
studied temperatures on several particles and processed to build this figure
are given in Supplementary fig. S4).\cite{SM} 
 The pre-peak “a” (Fig. \ref{figschematic}) comes from electron transitions
 from the O-$1s$ core level towards the lowest conduction band,\cite{Kikkawa14}
 which is
the Co-$3d$-$e_g$ band, where $e_g$ refers to the irreducible representation in the octahedral group of the Co octahedral environment. These
are antibonding hybridized Co-$3d$ - O-$2p$ states. 
The components “b” and “c” are due to transitions towards
higher conduction bands, which involve both the Li-$2s$ and
Co-$4s$ derived bands antibondng with O-$2p$ and are essentially
free-electron like bands.

According to Zhao \etal\cite{Zhao10} the
relative intensity of the pre-peak ``a'' to the ``b/c'' peaks
is connected to the degree of Co $3d$ - O $2p$ hybridization.
We thus studied this ratio of peak intensities as function of annealing
temperature, $T$, as shown in Fig.\ref{figOK}.
For annealed nanoflakes, in the range $T=20^\circ$C - 200$^\circ$C,
the ratio first increased, which would indicate an increase of the
Co-$3d$ – O-$2p$  hybridization for the 
$R\bar{3}m\rightarrow C2/m\rightarrow P2/m$ phase sequence.
On the other hand, for $T=200^\circ$C to $350^\circ$C  and
the transition $Fd\bar{3}m\rightarrow Fm\bar{3}m$, the $A(a)/A(bc)$ ratio 
decreased dramatically.

\begin{figure}[!htb]
  \includegraphics[width=\linewidth]{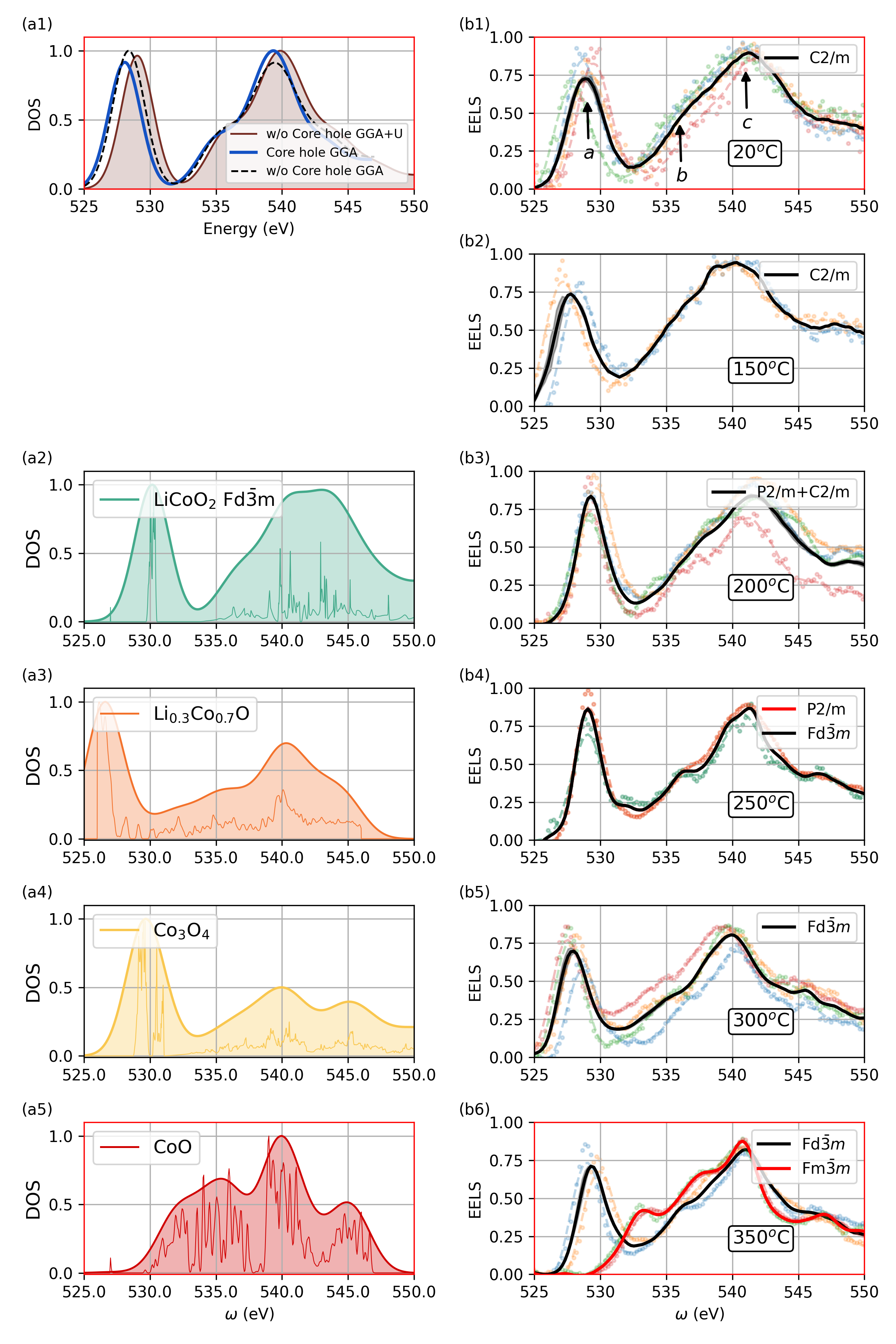} 
  \caption{(a) Simulated O-$K$ spectra for various
    systems: (a1) comparison of Eq.(\ref{eqOKsim}) including O-$1s$ core hole
    and matrix elements explicitly with simple O-$p$ PDOS both in GGA and with
    O-$p$ PDOS in GGA+U ($U=5$ eV) for LiCoO$_2$ in $R\bar{3}m$ structure,
    (a2-a5) O-$p$ PDOS in GGA+U for: (a2) LiCoO$_2$ in $Fd\bar{3}m$ model,
    (a3) Li$_{0.3}$Co$_{0.7}$O in $Fm\bar{3}m$ structure, (a4) Co$_3$O$_4$ in spinel $Fd\bar{3}m$ structure and (a5) CoO in rocksalt $Fm\bar{3}m$ structure. 
    (b) Experimental EELS for different samples (shown in scatter plots), with average data smoothened using Savitzky-Golay filter\cite{Savitzky64}
    shown in red/black for various temperatures\label{fig:ok_theory}} 
\end{figure}

Alternatively, we compare these spectra directly with calculated
O-$K$ spectra for various model systems in Fig. \ref{fig:ok_theory}.
First, in (a1) we compare different computational approaches for
LiCoO$_2$ in the $R\bar{3}m$ structure. This shows that the calculation
taking the core-hole and O-$1s$ to band states matrix elements explicitly
into account according to Eq. \ref{eqOKsim} is almost indistinguishable
from the simple O-$2p$ PDOS. Adding $U$ to the GGA just shifts the Co-$d$
bands up slightly but the spectra in either of these models agree well
with the experimental spectrum in (b1). The zero of energy in our calculation
is the highest occupied band (VBM)  but was then shifted by 527 eV to align the first ``a'' peak with experiment.  A Gaussian broadening by $\sim$1 eV is
applied to our calculated spectrum after removing the filled part of the
PDOS.
The unbroadened spectrum is shown in thinner lines underneath the shading. 
We expect negligible changes
in the closely related $C2/m$ and $P2/m$ phase. In the experimental
panels (b1-b6) the black thicker line is a smoothed average
over different samples. The sample details are provided in Supplemental
Information Fig. S4(a).\cite{SM}

Next, in panels (a2-a4) we show GGA+U results for the O-$p$ PDOS for
various models in an attempt to qualitatively understand changes in these
spectra. 
For the $Fd\bar{3}m$  calculations, we used a model provided by Materials
Project\cite{MP} with the above spacegroup. It contains both Li and Co
exclusively in octahedral sites.  One can see that 
there is little qualitative change between the $R\bar{3}m$ and $Fd\bar{3}m$
spectra, just as there is little change between the $P2/m$ and $Fd\bar{3}m$
spectrum in panel (b4). 
We also calculated a  $Fm\bar{3}m$ spectrum using a supercell of the
rocksalt structure with a random occupation of Li and Co atoms in a ratio of
$3/7$ Li/Co. Finally, we also considered spinel Co$_3$O$_4$  and
rocksalt CoO without any Li.

We note that in the rocksalt phase in GGA the $t_{2g}$ to $e_g$ crystal
field splitting in the octahedral environment is reduced compared to
that in $R\bar{3}m$ because of the larger Co-O distance. On the
other hand, the bands have significantly larger band dispersions, related
to the more 3D network type arrangement of Co and Li compared to the
layered structure in $R\bar{3}m$. Because of this we no longer have
a gap between separated $t_{2g}$ and $e_g$ bands but these two overlap
and lead to a metallic band structure, in disagreement with experiment.
This is why we need to use GGA+U to obtain a qualitatively meaningful
electronic structure. 
The gap in such systems arises from the formation of magnetic moments and
correlation effects in the partially filled Co-$3d$ derived bands.
For Co$_3$O$_4$ we obtain a ferromagnetic band structure in the primitive cell
when using GGA+U. For CoO we use an SQS model with random distribution of
up and down spins to model the paramagnetic state.  Likewise
for the Li containing $Fm\bar{3}m$  structure we use the SQS approach
to model both the Li/Co random  location and up and down spins on Co
sites. Band structures for these cases are shown in Supplemental Information
Figs. S7-S9.\cite{SM} Even with GGA+U, for the Li$_{0.3}$Co$_{0.7}$O
$Fm\bar{3}m$ structure we obtain a metallic band structure with the Fermi
level close to the top of the $d$ $t_{2g}$ band (Fig. S9). 

We can see that qualitatively all the simulated O-$K$ spectra stay
rather similar with similar  ``a'' and ``b/c'' peak relative
positions and intensity ratio.
The exception is CoO where the a peak is signficantly
less intense and closer to the b/c peaks and this agrees with the
experimental observation for the $Fm\bar{3}m$ spectra. 
From the entire series of experimental data, taking in to account the
sample to sample variation, the most obvious change
occurs for the $Fm\bar{3}m$ phase at 350$^\circ$C. From Fig. S4(a)
and Fig.\ref{fig:ok_theory}(b6), one can see 
that the strong drop in $A(a)/A(bc)$ is
related to a shift in the ``a'' peak to higher energies in qualitative
agreement with the CoO result in  Fig.\ref{fig:ok_theory}(a5).
This can also be seen in the simulations by Zhao \etal\cite{Zhao10}, with which our calculations agree quite well, even though their
calculation did not include the core-hole induced changes in the
density of states and was purely GGA.
The experimental spectra
for the $Fm\bar{3}m$ phase are much more resemblant of the pure CoO
phase than of the $Fm\bar{3}m$ Li$_{0.3}$Co$_{0.7}$O model we calculated.
This is an indication that at these higher temperatures, the system
may have  lost some  Li  from the rocksalt phase parts of the sample
by diffusion into other regions.

\subsection{Co-$L_{2,3}$ edge spectra}
\begin{figure}
  \includegraphics[width=8cm]{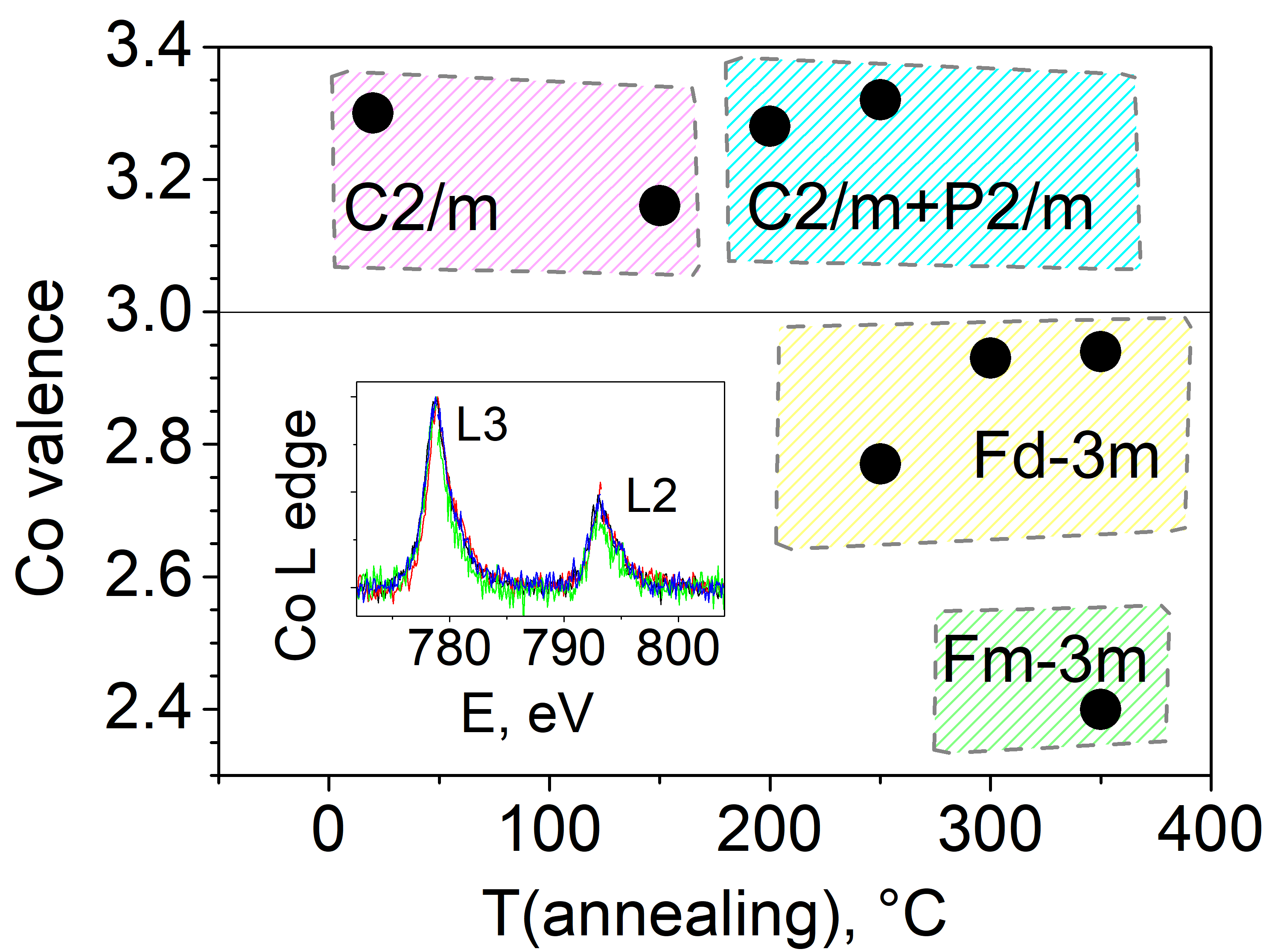}
  \caption{Average Co valence as deduced from the L$_3$/L$_2$2 edge peak ratio. \label{figCoL23}}
\end{figure}

The origin of the Co $L_3$,$L_2$ edge components is explained schematically
in  Fig. \ref{figschematic}.
They correpond to transitions from core levels Co $2p_{3/2}$ ($L_3$) and Co $2p_{1/2}$ ($L_2$). Both $L_3$ and $L_2$ include three components
each, marked  as ``a-c'' and ``d-f''.  Importantly, the  intensity ratio
$A(L_3)/A(L_2)$ in $3d$ transition metals is recognized as a parameter
which reflects the cation valence\cite{Zhao10,Wang2000,Egerton}. This is related
to many body effects  in these open shell states.  
Fig. \ref{figCoL23} presents the variation of Co valence, deduced from the calibration curve for $A(L_3)/A(L_2)$, presented in Supplementary Fig. S5.\cite{SM}
The step-like  changes of the Co valence in Fig.\ref{figCoL23}
are connected with the phase changes. 
Prior to annealing the mean Co valence of Li$_{0.37}$CoO$_2$
was around 3.3. A charge compensation of the Li loss
handled by Co only would give a valence
of 3.63. The lower experimental valence, together with the Co $3d$- O-$2p$ hybridization revealed by the O-$K$ pre-peak,
indicates a contribution from O-$2p$ holes in the charge compensation.
This is also consistent with findings by Wolverton and Zunger\cite{Wolverton98}
that within a sphere around Co, little change in charge density is
found upon Li loss but rather a change in Co-O bonding. 
At 150$^\circ$C, the Co valence is lowered to 3.15, probably
due to the increase of covalency and Co-O hybridizations in the growing $C2/m$
phase. The gradual growth of $P2/m$ domains induces an increase of the
mean Co valence. The $1:1$ ordering of the Li vacancies  drives the
charge ordering on Co sites, inducing Co atoms which donate to O more electronic charge than in a  random vacancy distribution.\cite{Miyoshi18,Kang2019}
Essentially, Co$^{4+}$ is formed near the Li vacancies. This is, probably, the reason why the Co valence is larger in $P2/m$ phase than in $C2/m$, although still with a degree of covalency in the Co-O bond. Finally, for $T>250^\circ$C, $Fd\bar{3}m$ and $Fm\bar{3}m$ phases have reduced average Co valence $<3+$, which evidences the increasing presence of reduced cations Co$^{2+}$. The pure
monoxide CoO has also the $Fm\bar{3}m$ rocksalt structure.\cite{Redman62} 
The reduction in average valence by the presence of Co$^{2+}$ in the
rocskalt phase is consistent with a loss of Li from these regions
found in the previous section from the O-$K$ spectra.

\subsection{Low energy EELS}

\begin{figure}
  \begin{center}
  \begin{subfigure}[b]{0.5\textwidth}
    \includegraphics[width=8cm]{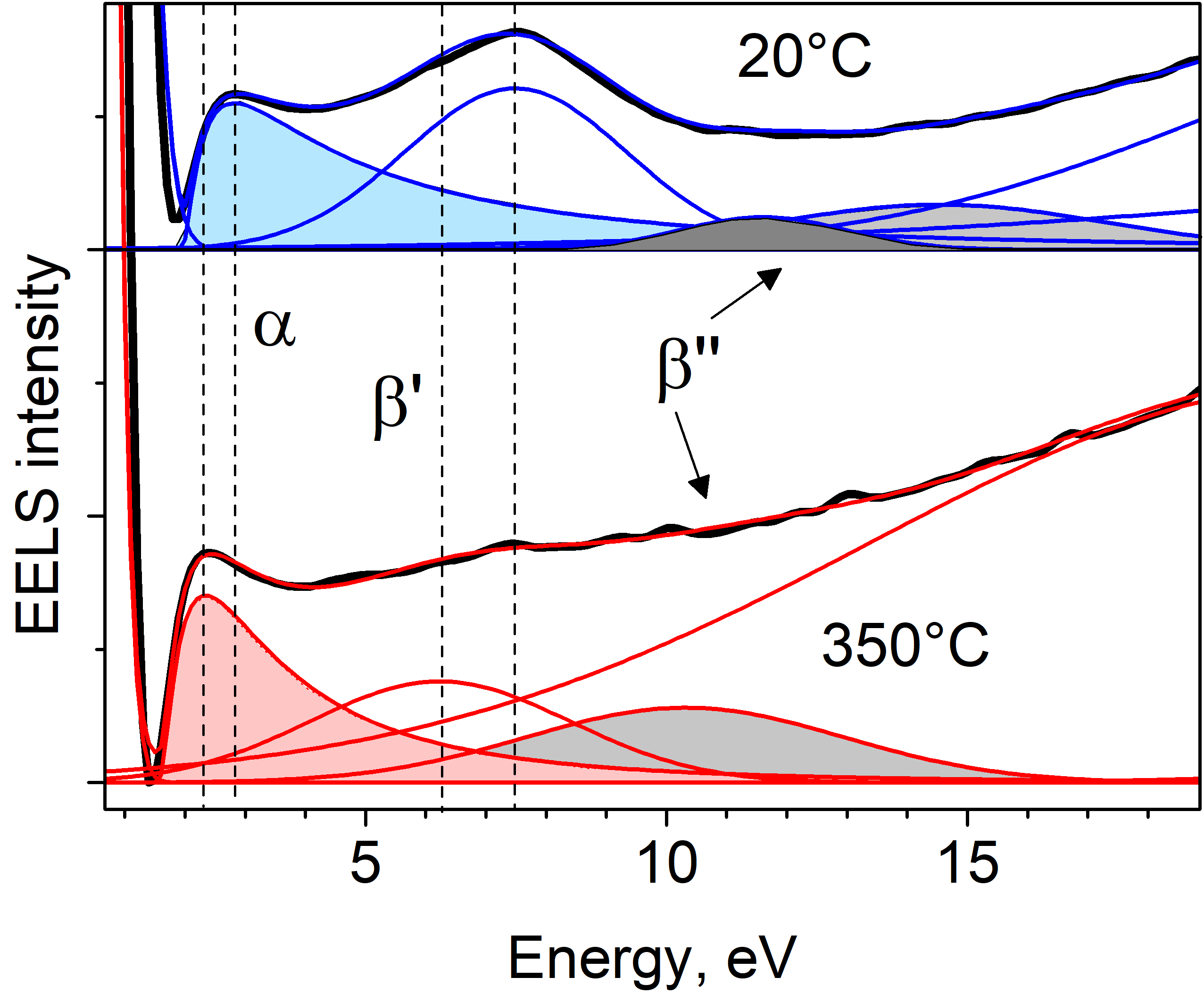}
\caption{}
  \end{subfigure}
  \begin{subfigure}[b]{0.5\textwidth}
    \includegraphics[width=8cm]{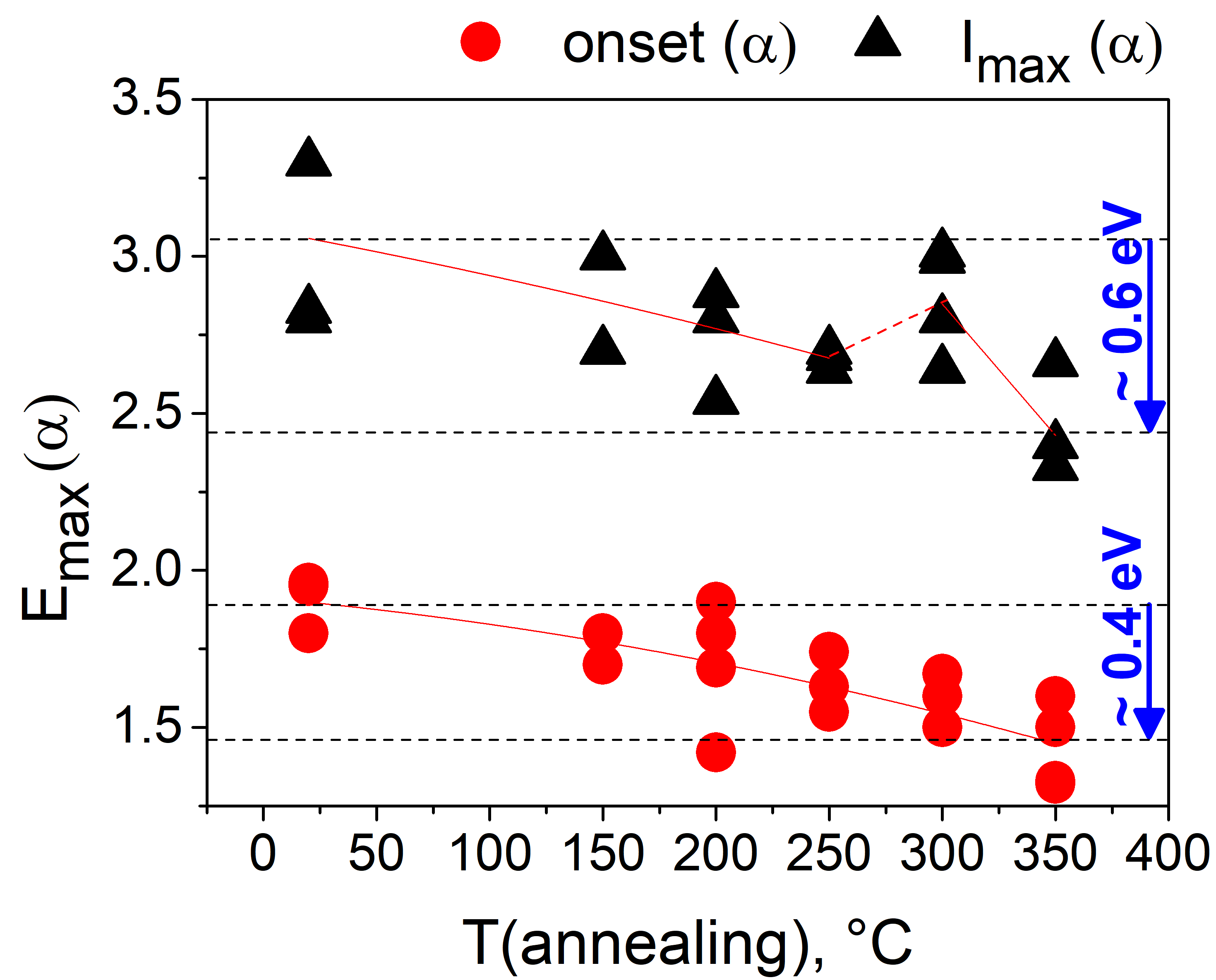}
    \caption{}
  \end{subfigure}
  \caption{Low energy loss spectra for near-bandgap electronic structure: (a) the fit model shown for two spectra after annealing at 20$^\circ$C and 350$^\circ$C, (b) onset of transition $\alpha$ and it's maximum plotted for all nanoflakes. \label{figEELS}}
  \end{center}
\end{figure}

\begin{figure}[!htb]
  \includegraphics[width=\linewidth]{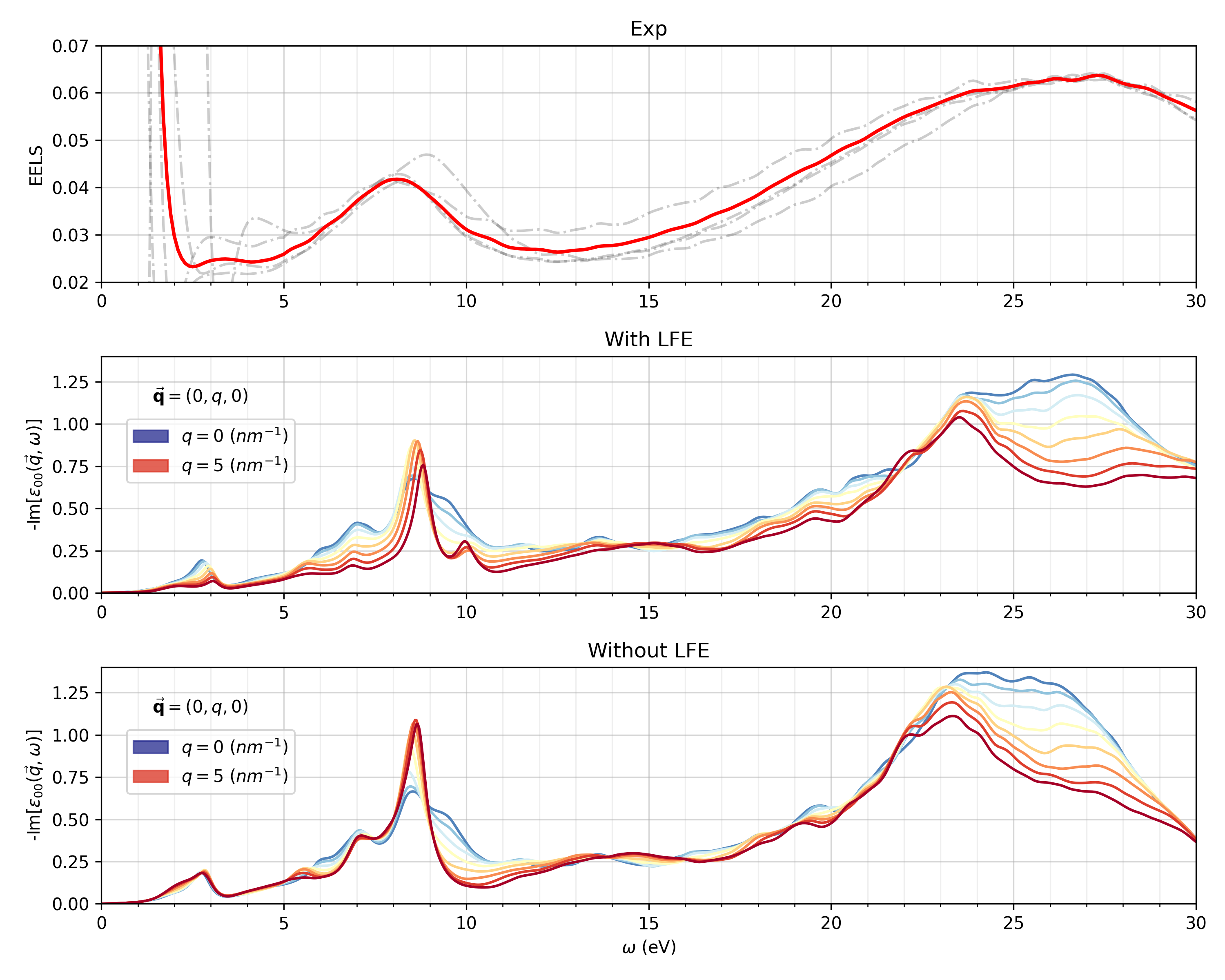} 
  \caption{EELS, from to to bottom: (a) experimental data at 20$^\circ$C
    the red curve is an average over three samples; 
    Calculated $\mathcal{A}_{\operatorname{ELSS}}(\mathbf{q}, \omega)$ at $\mathbf{q}=(0,q,0)$ for $q=0$(blue) to $q=0.5$(red) with (b) and 
    without(c)  Local Field effects. \label{fig:eels_theory}}. 
\end{figure}

Fig. \ref{figEELS} shows the measured low energy EELS for two temperatures.
The spectrum is deconvolved into an asymmetric $\alpha$ peak at about 3 eV and
rather broad $\beta$ peak at about 7.5 eV and a weak higher feature  labeled $\beta^{\prime\prime}$. As indicated in Fig. \ref{figschematic}  one
may loosely associate the $\alpha$ peak to transitions from the topmost 
Co-$t_{2g}$ part of the valence band to the $e_g$ conduction bands
and the $\beta$ peak to transitions from, a peak in density of states deeper
in the valence band with more O-$2p$ character to the same conduction band.
In between there is a region of low density of valence band states
which explains qualitatively why the $\alpha$-peak has a low density
tail toward higher energies and asymmetric shape. 
As shown in Supplementary Information (Fig. S10),\cite{SM} a simple convolution of
the occupied and empty densities of states, ignoring any momentum
conservation provides this type of general shape with an asymmetric
$\alpha$ and a broad $\beta$ peak.

However, a better approach is to compare these EELS spectra
with the calculated loss spectrum, 
which is shown in Fig. \ref{fig:eels_theory}.
As mentioned in the computational methods section, this corresponds to the
$q\approx0$ energy longitudinal response function. To be clear this
corresponds to vertical transitions but
integrated over the whole Brillouin zone. 
We here show calculations as function ${\bf q}$ and
both including and neglecting local field effects. The first weak peak around
2.5 eV agrees well with the experimental $\alpha$-peak. The next strong
peak centered at $\sim$8 eV agrees well with the experimental $\beta$-peak.
Even finer structure be recognized in the experiment as shoulder structures. 
The broad strong 
peak between 20-30 eV  corresponds to the plasmon. Using,
$\omega_P=\sqrt{4\pi ne^2/m_e}$  including 18 valence electrons, so not including the O-$2s$ electrons as valence, one obtains a plasmon energy of 27 eV,
while including the O-$2s$ would give 30 eV.
On the other hand the O-$2s$ derived bands lie about 20 eV below the VBM,
so  band-to-band transitions from these to the $e_g$ conduction band are also expected in this energy range, overlapping with the dominant plasmon peak. 

In this figure we show the spectrum as function {\bf q} up to about 5 nm$^{-1}$.
One may see that including local field effects, the $\alpha$ peak gradually
becomes smaller as {\bf q} increases but this effect is not seen when
local field effects are neglected.  In the present experiment, we estimate
that we integrate the spectrum near ${\bf q}=0$ up to about $q=5$ nm$^{-1}$.
To measure the spectrum as function of ${\bf q}$ one would need to
vary the central ${\bf q}$  away from the (000) spot in small steps and
within an even smaller  range or spot size $\delta q$.
Such measurements were done for the Li-$1s$ loss spectrum in Ref. \onlinecite{Kikkawa18}. 

Fig. \ref{figEELS} shows that both the onset and the location of the maximum  intensity of the $\alpha$-peak  shift to the lower energy on
increase of the annealing temperature.
Within the error of measurements, we observe a decrease
of these values by  $\sim$0.4 and  $\sim$0.6 eV respectively.
This decrease indicates a decrease of the $t_{2g}$-$e_g$ gap
and it agrees with the prediction\cite{Tarascon2004,Miedzinska1987}
of a band gap decrease with Co valence decrease. The decrease in valence
is also correlated with changes in Co-$3d$ - O-$2p$ hybridization
or degree of covalency as indicated by the O-$K$ spectra.

On the other hand, in CoO and Co$_3$O$_4$ the correlated
electronic structure can no longer be
explained purely within a standard DFT band-structure picture,\cite{vanElp}
although it is still possible to obtain a gap within DFT+U even for the
disordered paramagnetic phase using a polymorphous description, which
includes local symmetry breaking and spatial fluctuations.\cite{Trimarchi18}
Trimarchi \etal\cite{Trimarchi18}
obtain a gap of 2.25 eV for CoO in the paramagnetic
rocksalt structure. Our own calculations of the band structure within
GGA+U (given in the Supplemental Information S8)\cite{SM} give 
a gap of 1.94 in paramagnetic CoO
and this is indeed somewhat smaller 
than the gap obtained for LiCoO$_2$ in either the $R\bar{3}m$ (2.746 eV) or
$Fd\bar{3}m$ phases (2.710 eV).

From previous literature, the
prediction of decreasing gaps \cite{Tarascon2004,Miedzinska1987} is found to
hold for  Li$_x$CoO$_2$ with varying $x$. In the extreme case of $x=0$,
pure Co$_3$O$_4$ in the $Fd\bar{3}m$ phase also has  been reported
to have a smaller band gap (1.6 eV),\cite{Smart19}, which, however, is
related to a small polaron formation.
Note that this value of the gap is smaller than ours (1.724 eV) 
which does not include such polaronic effects. Whether or not such polaronic
effects are observed may depend on the time scale of the method with
which the gap is probed because the  polaron formation occurs at the time
scale of the atomic displacements. For example, one typically observes
a polaron in photoluminescence but not in optical absorption. The present
EELS measurements are closer to optical absorption. 

In the present case of Li$_{0.37}$CoO$_2$  this reduction of the gap
appears to be caused by the structural transitions, and changes of cation order.
It is well known that disorder may reduce the band gap by forming
defect like band tails in the gap near the band edges. However,
in the present case, the situation may be more complex by the increasing
importance of not only structural and cation disorder fluctuations but
also magnetic moment formation and fluctuations in magnetic moment orientation
and $d$-band correlation effects, including possibly strong electron-phonon
coupling as occurs in self-trapped polaron formation. Disentangling these
various effects is beyond the scope of the present paper. 

\subsection{Conductivity}
\begin{figure}
  \includegraphics[width=8cm]{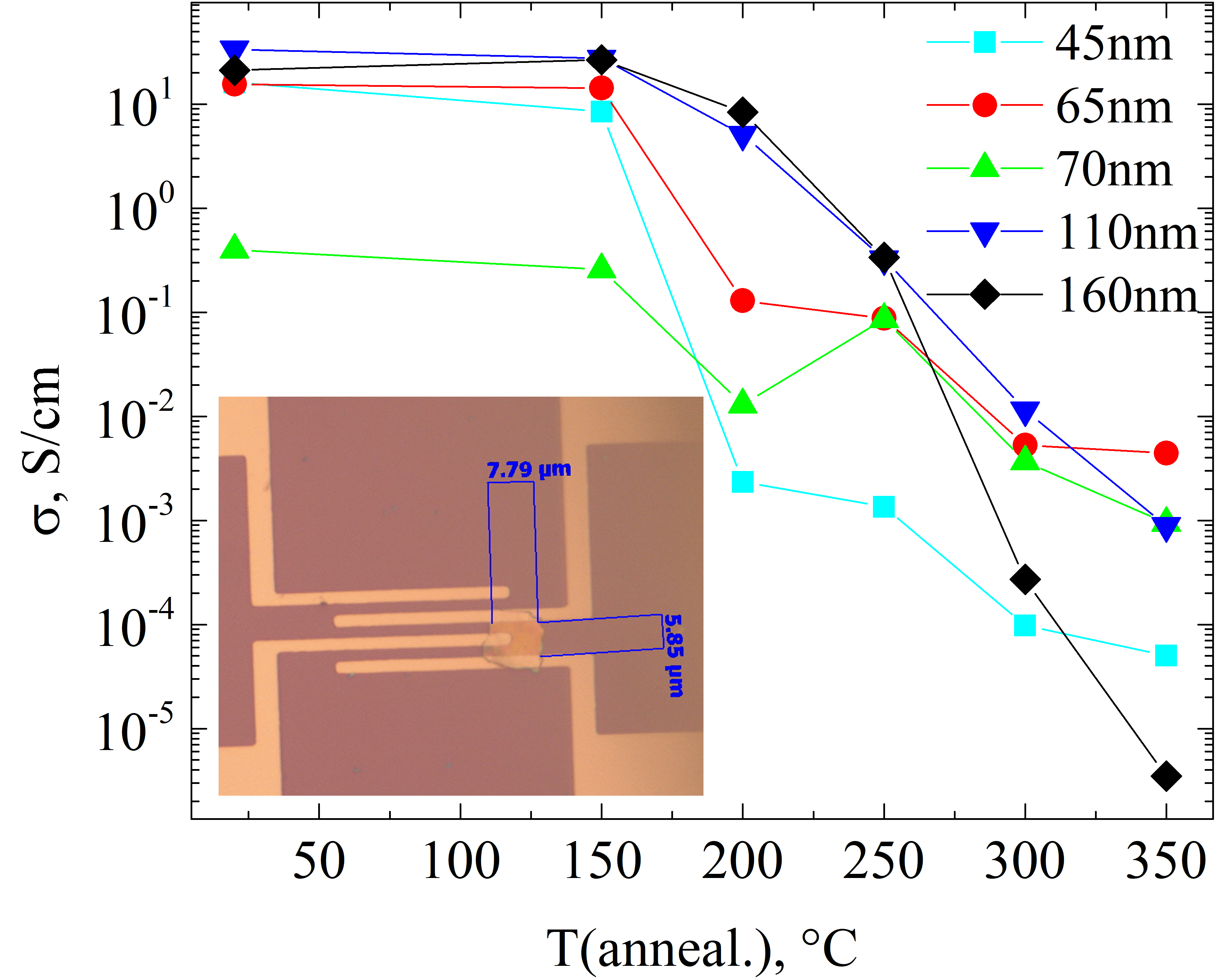}
  \caption{Electrical conductivity as function of annealing temperature.
    Inset: device structure showing electrical contacts. \label{figcond} }
\end{figure}
Finally, we examine how the conductivity changes with annealing temperature
and whether it is related to the observed band gap decrease and changes
in valence, hybridization evidenced by the spectroscopic investigations
of the previous sections.  The conductivity changes as function of
annealing temperature are shown in Fig. \ref{figcond}. 
The conductivity remains at its highest value for $T\le150^\circ$C corresponding to the 
$C2/m$ phase. A decrease in conductivity is
observed for $T\ge 150^\circ$C. 
At 200$^\circ$C it corresponds to a macroscopic $P2/m$ matrix,
possibly with the beginning of  $Fd\bar{3}m$ nanodomains at 200$^\circ$C (Supplementary Fig. S1(3)).\cite{SM}
This $Fd\bar{3}m$ phase with its lower covalency and O-$2p$ – Co-$3d$
hybridization at the conduction band edges is, probably, at the origin
of the start of  conductivity decrease,  as charge carriers tend to become more localized for less hybridized levels. This decrease is
stronger for the macroscopic
$Fd\bar{3}m$ and $Fm\bar{3}m$  phases at 250-350$^\circ$C.
The spinel structure for $x=2/3$ was reported as insulating in STM when cation migration forces random occupation of sites.\cite{Miyoshi18},
In our case, the spinel type phase with $x\approx1/3$  is also found to
be insulating.

On the other hand, as mentioned
earlier, there are also indications that on heating the Li concentration
may be decreasing in the sample due to out-diffusion. The poorer the system
is in Li, the more correlated the electronic structure becomes with
increasing polaronic effects\cite{Smart19} and
this may also be part of the reason for the conductivity decrease.
While for Li$_x$CoO$_2$ in the $R\bar{3}m$ and the closely related
$C2/m$ and $P2/m$ structures,  a normal band picture
still holds with some $p$-type hole doping due to $x<1$, in the $Fd\bar{3}m$
and even more so, in the $Fm\bar{3}m$ phases, the starting picture
of a band insulator becomes untenable because of the overlapping of the
$t_{2g}$ and $e_{g}$ bands. In the $Fd\bar{3}m$ one would even have
inverted $t_{2g}$ and $e_g$ levels at tetrahedral Co-sites. The origin
of the band gap in this case does no longer arise from a well separated
nearly filled $t_{2g}$ band below an empty $e_g$ band with low-spin
configuration, but from correlation effects leading to magnetic moments
and a gap forming between majority and minority spin electrons.
The nature of the conductivity is thus clearly expected to be dramatically
changing once these phases come into play. 
The inter-layer cation mixing might thus be considered to
be detrimental for the conductivity.

\section{Conclusions}
To conclude, we have presented in this paper a comprehensive
study  of the structural changes, electronic properties as derived from
electron energy loss spectroscopy correlated with first-principles calculations,  and conductivity measurements of Li$_{0.37}$CoO$_2$ nanoflakes subjected to
heating.
The O-$K$ spectra were interpreted in terms of O-$p$ PDOS modulated by
the dipole matrix elements linking the conduction states to the
O-$1s$ core-wavefunction in an all-electron approach
and taking the presence of the core-hole into account.
The low energy loss spectra were well described by the calculated
loss function $-\mathrm{Im}[\varepsilon^{-1}({\bf q},\omega)]$ for small
  {\bf q}. Predictions are made for the ${\bf q}$-dependence of these spectra.
  They offer a basic interpretation of the main features, whose trends
  are studied upon heating. 
  
The Li$_{0.37}$CoO$_2$ flakes are initially (between 150$^\circ$ C and 200$^\circ$ C annealing temperature) found to experience phase transitions
from the rhombohedral $R\bar{3}m$ phase to the monoclinic phases $C2/m$ and $P2/m$ with disordered and
ordered, Li vacancies respectively.  During these transitions,
the Co valence increases in parallel with increasing Co-$3d$ – O-$2p$
hybridization as evidenced from the interpretation of the peak ratio
changes in O-$K$ and Co-$L_{2,3}$ spectra. 
Upon further heating above 250$^\circ$C and completed by 350$^\circ$C
when partial or full Li-Co interlayer mixing
happens in the spinel-type $Fd\bar{3}m$ and rocksalt-type $Fm\bar{3}m$ phases,
the Co nominal valency decreases, as well as the hybridization of Co-$3d$ – O-$2p$ states. A band gap decrease is observed when these phases start to form.
The increasing presence of Co$^{2+}$ indicated by $L_{2,3}$ spectra
is consistent with a loss of Li from the
rocksalt phase regions, as also indicated by the O-$K$ spectra. 
These changes are related to the dramatically modified band structure which
is no longer in a low-spin state. Band structure calculations
at the GGA level indicate a significant overlap of $t_{2g}$ and $e_{g}$ bands in the $Fm\bar{3}m$ phase with a metallic band structure, which becomes unstable
toward magnetic moment formation. The latter can be described within the
DFT+U methodology and allowing the magnetic moments to occur in a disordered
manner. 
This leads to a three to six orders of magnitude decrease in  conductivity at temperatures where
these phases start to form because of increased Li and Co interdiffusion
forming a 3D network instead of a layered phase.

\acknowledgements{ This work was supported by the U.S. Air Force Office
  of Scientific Research under Grant No. FA9550-18-1-0030.  The calculations made use of the High Performance Computing Resource in the Core Facility for Advanced Research Computing at Case Western Reserve University.}

\bibliography{lmto,dft,gpaw,lco}
\end{document}